\documentclass[%
 reprint,
 superscriptaddress,
 amsmath,amssymb,
 prl,
]{revtex4-2}

\usepackage{graphicx}% Include figure files
\usepackage[caption=false]{subfig}
\usepackage{dcolumn}% Align table columns on decimal point
\usepackage{bm}% bold math
\usepackage{hyperref}% add hypertext capabilities

\usepackage{siunitx}

\begin{document}

\title{Stability of the Modulator in a Plasma-Modulated Plasma Accelerator}

\author{J. J. van de Wetering}
\email{johannes.vandewetering@physics.ox.ac.uk}
\affiliation{John Adams Institute for Accelerator Science and Department of Physics, University of Oxford, Denys Wilkinson Building, Keble Road, Oxford OX1 3RH, United Kingdom}%
\author{S. M. Hooker}%
\affiliation{John Adams Institute for Accelerator Science and Department of Physics, University of Oxford, Denys Wilkinson Building, Keble Road, Oxford OX1 3RH, United Kingdom}%
\author{R. Walczak}%
\affiliation{John Adams Institute for Accelerator Science and Department of Physics, University of Oxford, Denys Wilkinson Building, Keble Road, Oxford OX1 3RH, United Kingdom}%
\affiliation{Somerville College, Woodstock Road, Oxford OX2 6HD, United  Kingdom}%

\date{\today}% It is always \today, today,
             %  but any date may be explicitly specified

\begin{abstract}
We explore the regime of operation of the modulator stage of a recently proposed laser-plasma accelerator scheme [Phys. Rev. Lett. \textbf{127}, 184801 (2021)], dubbed the Plasma-Modulated Plasma Accelerator (P-MoPA). The P-MoPA scheme offers a potential route to high-repetition-rate, GeV-scale plasma accelerators driven by picosecond-duration laser pulses from, for example, kilohertz thin-disk lasers. The first stage of the P-MoPA scheme is a plasma modulator in which a long, high-energy `drive' pulse is spectrally modulated by co-propagating in a plasma channel with the low-amplitude plasma wave driven by a short, low-energy `seed' pulse. The spectrally modulated drive pulse is converted to a train of short pulses, by introducing dispersion, which can resonantly drive a large wakefield in a subsequent accelerator stage with the same on-axis plasma density as the modulator. In this paper we derive the 3D analytic theory for the evolution of the drive pulse in the plasma modulator and show that the spectral modulation is independent of transverse coordinate, which is ideal for compression into a pulse train. We then identify a transverse mode instability (TMI), similar to the TMI observed in optical fiber lasers, which sets limits on the energy of the drive pulse for a given set of laser-plasma parameters. We compare this analytic theory with particle-in-cell (PIC) simulations and find that even higher energy drive pulses can be modulated than those demonstrated in the original proposal.
\end{abstract}

%\keywords{Suggested keywords}%Use showkeys class option if keyword
                              %display desired
\maketitle

\section{Introduction}

In a laser-plasma accelerator (LPA), plasma oscillations are driven by pushing free electrons away from an ultrashort laser pulse via the ponderomotive force. The heavier ions remain approximately stationary relative to the electrons, thus the electron-ion charge separation collectively forms a strong electric field which can be used to accelerate charged particles. The plasma wave driven in this way will have a phase velocity set by the group velocity of the laser pulse, which is well suited for the acceleration of relativistic charged particles. The accelerating gradients achievable by LPA are set by the wavebreaking field $E_0=m_e\omega_pc/e$ and can be on the order of 100 GV/m \cite{PhysRevLett.43.267}, three orders of magnitude larger than those possible in radio-frequency cavities. Here the plasma frequency $\omega_p = (n_e e^2 / m_e \epsilon_0)^{1/2}$, where $n_e$ is the electron density. 

Efficient excitation of the plasma wave by a single laser pulse requires that the duration of the pulse is less than half the plasma period $T_p = 2 \pi / \omega_p$. For plasma densities of interest $T_p$ is in the 100 fs range, and hence single-pulse LPAs first became practical with the development of chirped pulse amplification (CPA) \cite{STRICKLAND1985219}, which allowed joule-scale pulses to be compressed to sub-picosecond durations. Ever since, most experimental demonstrations of LPAs have used high intensity ultrashort laser pulses from Ti:sapphire CPA laser systems. However, these systems suffer from low ($\sim$ 0.1-10 Hz) repetition rates \cite{doi:10.1063/1.4773687} and poor ($<0.1\%$) electrical-to-optical energy efficiencies \cite{PWASC}. Despite the advantages gained by being much more compact, the low efficiency and repetition rate of the laser drivers used today severely limit the number of applications for which LPAs offer an advantage over conventional, radio-frequency particle accelerators.

It is important, therefore, to consider alternative laser systems and/or develop novel approaches for driving LPAs. Contemporary thin-disk lasers are efficient and can already provide joule-scale pulses at kHz repetition rates \cite{Herkommer:20,Nagel:21,Produit:21}. However, they cannot drive a LPA directly since the small bandwidth of their gain media limits the duration of the pulses they generate to \cite{Paschotta2001,Sudmeyer2009,Baer:10} $\tau \gtrsim \SI{1}{ps}$, which is much longer than the plasma period. We note that pulses from thin-disk lasers have been compressed to a duration below \SI{100}{fs}, following spectral broadening via self-phase modulation in a gas \cite{Kaumanns:21}. However, to date this approach has been limited to pulse energies below \SI{120}{mJ}.

\begin{figure*}
    \centering
    \includegraphics[width=\linewidth]{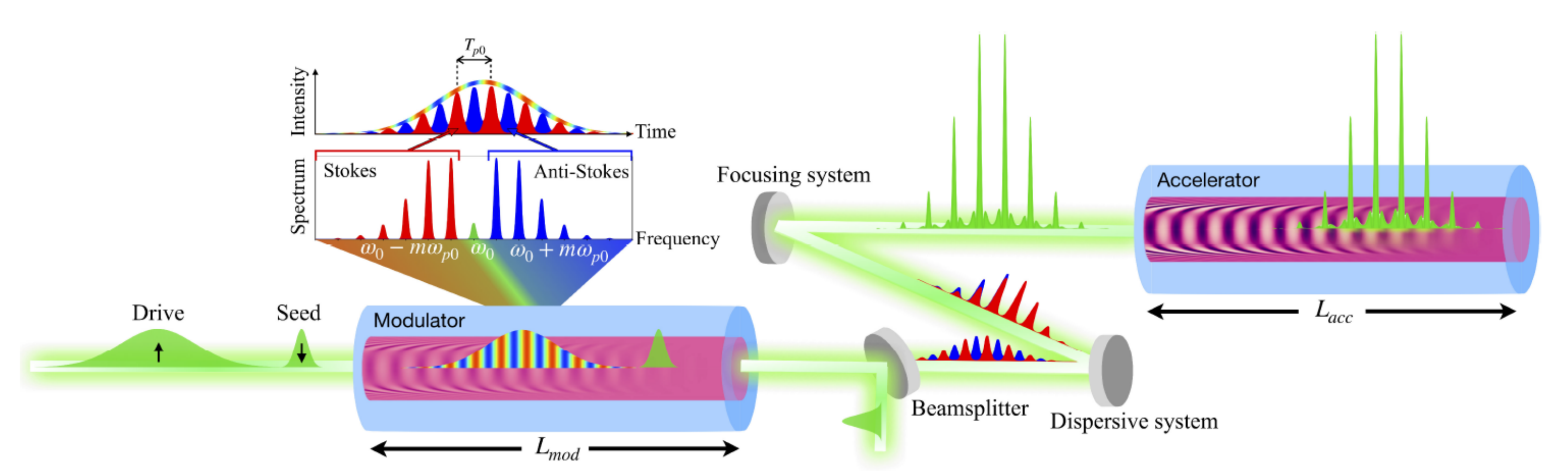}
    \caption[Outline of the P-MoPA Concept]{[Color online]. Outline of the P-MoPA scheme from the original proposal \cite{PhysRevLett.127.184801}. A short, low energy seed pulse excites a small wake in the modulator stage which spectrally modulates a long, high energy drive pulse into interleaving redshifted (Stokes) and blueshifted (anti-Stokes) pulse trains whilst maintaining a smooth envelope. Chromatic dispersion is then applied to the spectrally modulated drive pulse to compress it into a multipulse train, which can then be used to resonantly drive a wakefield in the accelerator stage with the same density as the modulator.}
    \label{fig:PhysRevLett.127.184801}
\end{figure*}

With the objective of applying the desirable features of thin-disk lasers to LPAs, some of the present authors recently proposed a scheme, illustrated in Fig.\ \ref{fig:PhysRevLett.127.184801}, for converting picosecond-duration pulses to a train of shorter pulses that could be used to resonantly excite a plasma wave \cite{PhysRevLett.127.184801}. In this scheme, which we call the Plasma-Modulated Plasma Accelerator (P-MoPA), a high-energy, picosecond-duration `drive' pulse is modulated spectrally by co-propagating it in a plasma channel with a low amplitude ($\sim 1\%$) plasma wave driven by a low-energy, short `seed' pulse. To first order, the spectral modulation takes the form of a set of sidebands of angular frequencies $\omega_m = \omega_L + m \omega_{p0}$, where $\omega_L$ is the angular frequency of the input drive pulse, $m = \pm1, \pm2, \pm3, \ldots$ is the sideband order, and $\omega_{p0}$ is the plasma frequency on the axis of the plasma channel. The spectrally-modulated drive pulse is converted into a temporal modulation by passing it through a dispersive optical system that removes the relative spectral phase, $\psi_m = - |m| \pi / 2$, of each sideband. This forms a train of short pulses, spaced temporally by $T_{p0}$, which can resonantly drive a large amplitude plasma wave in a plasma accelerator stage with the same axial density as the modulator.

In our earlier work a one-dimensional (1D) fluid model, and 2D particle-in-cell (PIC) simulations were used to demonstrate the operation of the plasma modulator and accelerator stages, and to show that GeV-scale energy gains could be obtained from existing thin-disk laser systems. In this paper we derive a full 3D theory of seeded spectral modulation and we use this to establish the useful operating regime for the modulator stage in the P-MoPA. We find that the range of operation of the modulator is determined by the onset of the transverse mode instability (TMI), similar to the TMI observed in high power fiber laser systems \cite{Eidam:10,Eidam:11,Jauregui:11,Smith:11,Jauregui:20}. This analysis is used to establish the regime of parameter space for which the modulator can be operated successfully. The results of the 3D analytic theory are compared with particle-in-cell (PIC) simulations, and are found to be in good agreement. We find that even higher energy drive pulses can be modulated than those considered in the original proposal.

\section{The Plasma Modulator}

\subsection{Seeded Spectral Modulation in Plasma Channels}

Propagation of the envelope of a laser pulse in an axisymmetric plasma channel of electron density $n_0(r)=n_{00}+\delta n_0(r)$ with a small wake $\delta n(r,\xi;|a|^2)$ can be approximately described by the paraxial wave equation \cite{doi:10.1063/1.860707,NEAndreev_1994,doi:10.1063/1.872134,doi:10.1063/1.3691837} (see Supplemental Material \cite{supp} for its derivation)
\begin{align}\label{eq:GNLS}
    &\left[\frac{i}{\omega_L}\frac{\partial}{\partial\tau} + \frac{c^2}{2\omega_L^2}\Delta_\perp\right] a = \nonumber \\ &\frac{\omega_p^2}{2\omega_L^2n_0}\left[\delta n_0(r) + \delta n(r,\xi;|a|^2) - n_0(r)|a|^2/4\right]a
\end{align}
where $a(r,\theta,\xi,\tau)$ is the envelope of the pulse's normalized vector potential, $\omega_L$ is the laser frequency and the propagation is described in co-moving coordinates $\xi=z-v_gt$, $\tau=t$, with $v_g/c = (1-\omega_{p0}^2/\omega_L^2)^{1/2}$ defined as the group velocity of electromagnetic plane waves in uniform plasma of density $n_{00}$, corresponding to the on-axis plasma channel frequency $\omega_{p0}=\omega_p(r=0)$. This group velocity may differ from the group velocity of a tightly focused laser pulse \cite{Esarey:95,PhysRevE.59.1082}. Nonlinearities come from weakly relativistic effects as well as from interaction between the pulse and its own excited wake.

\begin{figure}[tb]
    \centering
        \includegraphics[width=\linewidth]{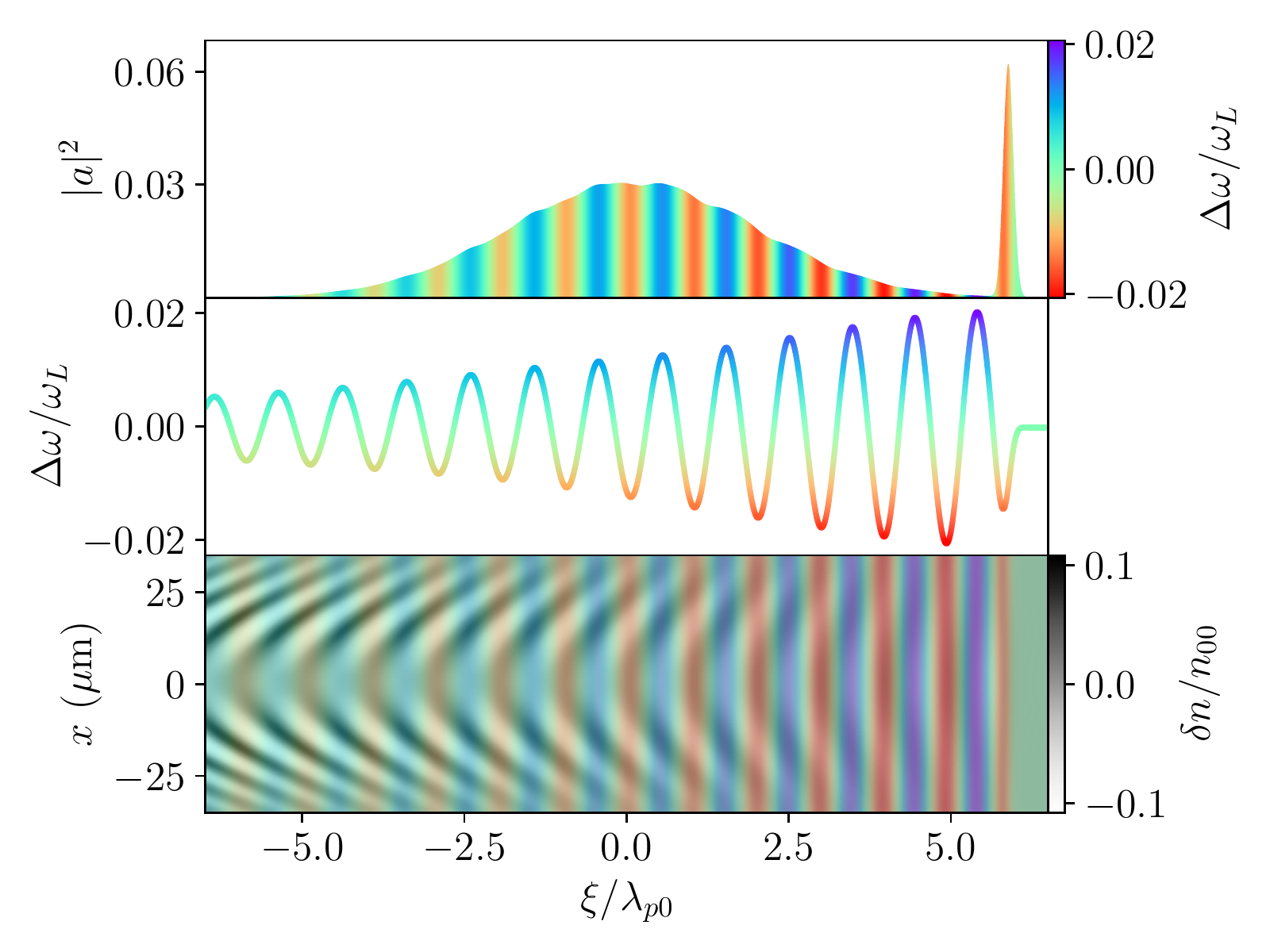}
    \caption[Radially independent spectral modulation]{[Color online]. 2D PIC simulation of the modulator stage in a matched parabolic plasma channel of matched spot size $w_0=\SI{30}{\micro m}$ with $W_\text{seed}=\SI{50}{mJ}$ and $W_\text{drive}=\SI{0.6}{J}$, $\tau_\text{drive}=\SI{1}{ps}$. The top panel shows the on-axis longitudinal intensity profiles $|a|^2$ for the seed and drive pulses. The middle panel plots the on-axis instantaneous frequency calculated by a Hilbert transform. The bottom panel displays the full 2D distributions of the relative amplitude $\delta n / n_{00}$ of the plasma wave and the relative frequency modulation $\Delta \omega / \omega_L$.}
    \label{fig:specmodsimfull}
\end{figure}

Successful seeded spectral modulation requires relativistic and self-wake effects to be negligible, which reduces Eq. (\ref{eq:GNLS}) to a linear paraxial wave equation
\begin{align}\label{eq:seedparaxial}
    \left[\frac{i}{\omega_L}\frac{\partial}{\partial\tau} + \frac{c^2}{2\omega_L^2}\Delta_\perp\right] a &= \frac{\omega_p^2}{2\omega_L^2n_0}\left[\delta n_0(r) + \delta n(r,\xi;|a_s|^2)\right]a
\end{align}
where $a_s$ denotes the seed pulse envelope, whose intensity we assume to be unchanging relative to the modulating drive pulse envelope $a$. We also demand that the channel is matched to the spot size $w_0$ of the seed and drive pulses. A matched parabolic channel and its respective unperturbed Gaussian drive pulse envelope take the form \cite{1126872, FIRTH1977226, PhysRevE.59.1082}
\begin{align}\label{eq:unperturbed}
    &n_0(r) = n_{00} + \Delta n(r/w_0)^2 \nonumber\\
    &a(r,\xi,\tau) = a_0f(\xi)\exp\left(-\frac{r^2}{w_0^2}-i\omega_L\tau\frac{2c^2}{\omega_L^2w_0^2}\right)
\end{align}
where $\Delta n\equiv(\pi r_ew_0^2)^{-1}$ is the channel depth parameter, $r_e$ is the classical electron radius and $0\leq f(\xi)\leq 1$ is the slowly-varying longitudinal envelope of the drive pulse. Assuming the seed wake is small relative to the channel depth parameter $|\delta n(r,\xi;|a_s|^2)|\ll\Delta n$, we can apply time-independent perturbation theory to Eq. (\ref{eq:seedparaxial}), yielding the following modulation to the total phase of the laser pulse
\begin{align}\label{eq:E1}
    \Phi(\xi,\tau) &= k_L\xi-\frac{2c^2\tau}{\omega_Lw_0^2}\left(1+\left\langle\frac{\delta n(r,\xi;|a_s|^2)}{\Delta n}\right\rangle_\perp\right)
\end{align}
where $k_L=\eta\omega_L/c$ is the laser wavenumber with the on-axis plasma index of refraction $\eta=(1-\omega_{p0}^2/\omega_L^2)^{1/2}$ and $\langle(\ldots)\rangle_\perp=(4/w_0^2)\int_0^\infty(\ldots)\exp(-2r^2/w_0^2)rdr$ denotes the intensity-weighted transverse average. Using this expression we can also retrieve the shift in instantaneous frequency
\begin{align}\label{eq:specshift}
    \frac{\Delta\omega(\xi;L_\text{mod})}{\omega_L} &= - L_\text{mod}\frac{2c^2}{\omega_L^2w_0^2}\left\langle\frac{\partial}{\partial\xi}\frac{\delta n(r,\xi;|a_s|^2)}{\Delta n}\right\rangle_\perp
\end{align}
where $L_\text{mod}=v_g\tau$ is the modulator length. This predicts that the spectral modulation amounts to a radial averaging of the longitudinal gradient of the wake weighted by the transverse intensity profile of the drive pulse. This independence of the spectral modulation of radial position is desirable as it allows the spectral modulation to be converted into a temporal one by applying the same chromatic dispersion across the entire cross-section of the modulated pulse. Figure \ref{fig:specmodsimfull} shows the results of a 2D PIC simulation which demonstrates that the spectral modulation is indeed independent of the transverse coordinate, despite the fact that the amplitude and phase of the plasma wave \emph{does} depend on the radial coordinate, as predicted by Eq. (\ref{eq:specshift}). Figure \ref{fig:specmodsimfull} also shows suppression of the spectral modulation towards the tail of the drive pulse, which is most apparent in the plot of the retrieved instantaneous frequency. This is also expected from Eq.\ \eqref{eq:specshift}, since the large curvature of the wavefronts of the plasma wave towards the tail of the drive pulse leads to a reduction in the spectral modulation when the longitudinal gradient of the wave is averaged radially. Note that in the Supplemental Material \cite{supp} we compare the results of PIC simulations in 2D and cylindrical geometry. All PIC simulations presented in this paper were performed with axial plasma density $n_{00}=\SI{2.5e17}{cm^{-3}}$, seed and drive laser wavelength $\lambda_L=\SI{1030}{nm}$, seed pulse FWHM duration $\tau_\text{seed}=\SI{40}{fs}$ and modulator length $L_\text{mod}=\SI{110}{mm}$. A complete list of simulation parameters is included in the Supplemental Material \cite{supp}. 

\subsection{Channel Suppression of Spectral Modulation}

As observed in Fig.\ \ref{fig:specmodsimfull}, Eq. (\ref{eq:specshift}) implies that the spectral modulation of the drive pulse is limited by wave-front curvature of the plasma wave. For low amplitude wakes this curvature is dominated by the transverse profile of the plasma channel. For square channels the wake has flat phase fronts over most of the transverse profile of the lowest-order mode of the channel. However, for channels with a curved transverse profile, such as parabolic channels, the wave-fronts of the plasma wave are curved, which can strongly suppress the spectral modulation. 

\begin{figure}
    \centering
    \includegraphics[width=\linewidth]{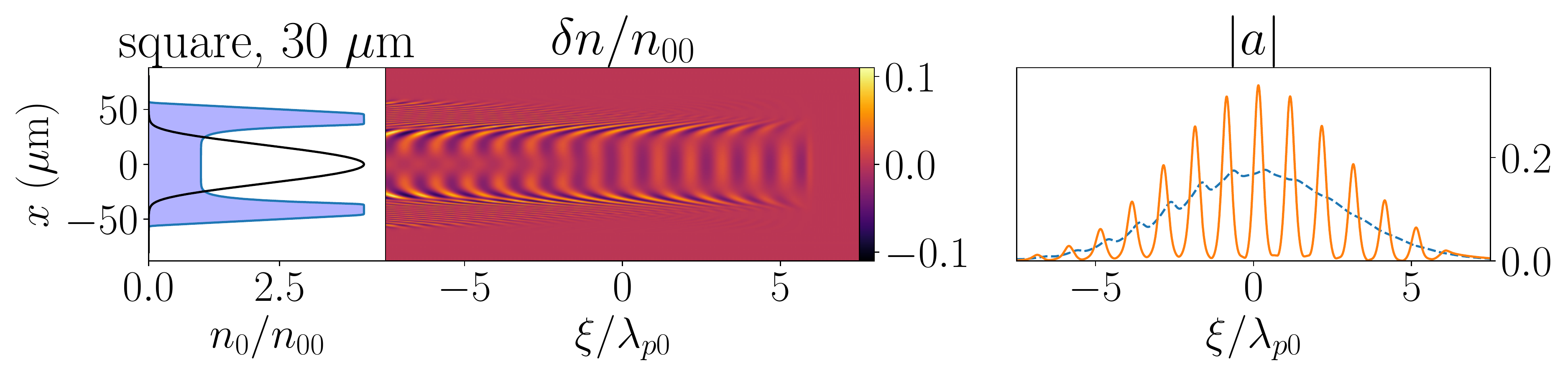}
    \includegraphics[width=\linewidth]{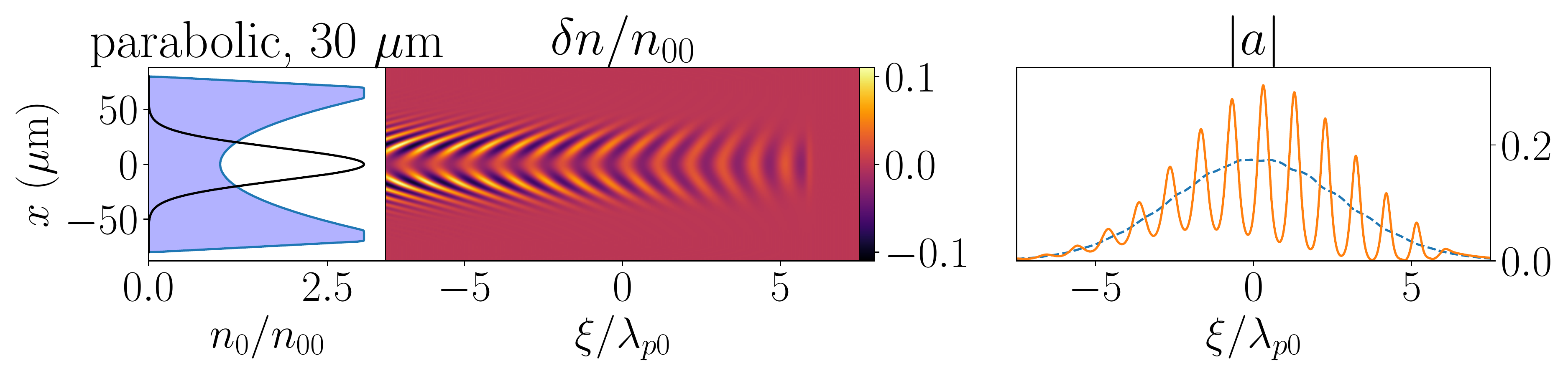}
    \includegraphics[width=\linewidth]{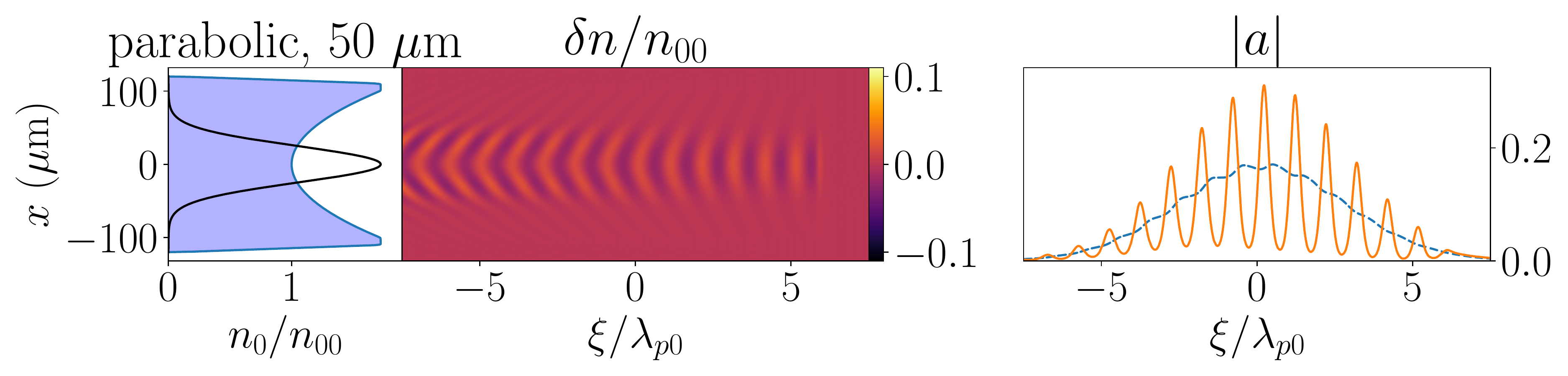}
    \caption{[Color online]. Comparison of the performance of modulators with plasma channels of different transverse profiles each with wall thicknesses of $\SI{20}{\micro m}$ (see Supplemental Material for their parameterizations \cite{supp}). The left panels show the transverse electron density (blue) and normalized guided intensity (black) profiles of the channels for: top, a square channel of diameter \SI{30}{\micro m}; middle, a parabolic channel of matched spot size \SI{30}{\micro m}; bottom, a parabolic channel of matched spot size \SI{50}{\micro m} (with $(50/30)^2\times$ more seed and drive pulse energy to account for the larger spot size). The middle panels show the relative wake amplitudes $\delta n / n_{00}$ at the end of the modulator, calculated by 2D PIC simulations. The right panels show the on-axis pulse envelopes $|a|$ at the end of the modulator before (dashed blue) and after (solid orange) the expected \cite{PhysRevLett.127.184801_supp} sideband spectral phase $\psi_m$ was removed (see Supplemental Material \cite{supp}).}
    \label{fig:specmodsupp2}
\end{figure}

Figure \ref{fig:specmodsupp2} shows the results of 2D PIC simulations that compare the performance of a modulator with square and parabolic plasma channels. It can be seen that for the square channel the wave-fronts of the plasma wave are flat across most of the channel width, and as a consequence the pulse train generated by removal of the sideband spectral phase exhibits strong temporal modulation over the entire duration of the pulse train. In contrast, the wave-fronts of the wake driven in a parabolic channel of \SI{30}{\micro m} matched spot size are strongly curved, and this curvature increases towards the tail of the drive pulse. As a consequence, the generated pulse train does not exhibit complete intensity modulation near its tail, which would reduce the amplitude of the plasma wave it could drive in the accelerator stage. As shown in Fig.\ \ref{fig:specmodsupp2}, increasing the matched spot size of the parabolic channel to \SI{50}{\micro m} reduces the wake curvature, which leads to improved modulation of the generated pulse train. We note that the deleterious effects of wave-front curvature would be even more pronounced in 3D geometry (see Supplemental Material \cite{supp}).

\section{Stability of the Plasma Modulator}

It is important to understand the extent to which instabilities will arise in the modulator, and the range of laser and plasma parameters for which any deleterious effects arising from them can be avoided. Since we do not want to waste any of the drive pulse energy within the modulator stage, we would like its envelope to remain smooth as it propagates. Although the specific set of parameters used in the original P-MoPA proposal \cite{PhysRevLett.127.184801} were shown to work in simulation, for the scheme to be practical we want a large, well-defined parameter space where it is stable.

\begin{figure*}[tb]
    \centering
        \hspace*{-1.0em}
        \subfloat[\label{subfig:a} $\tau_\text{drive}=\SI{1}{ps}$,\, $W_\text{drive}=\SI{0.6}{J}$]{
            \includegraphics[width=0.333\linewidth]{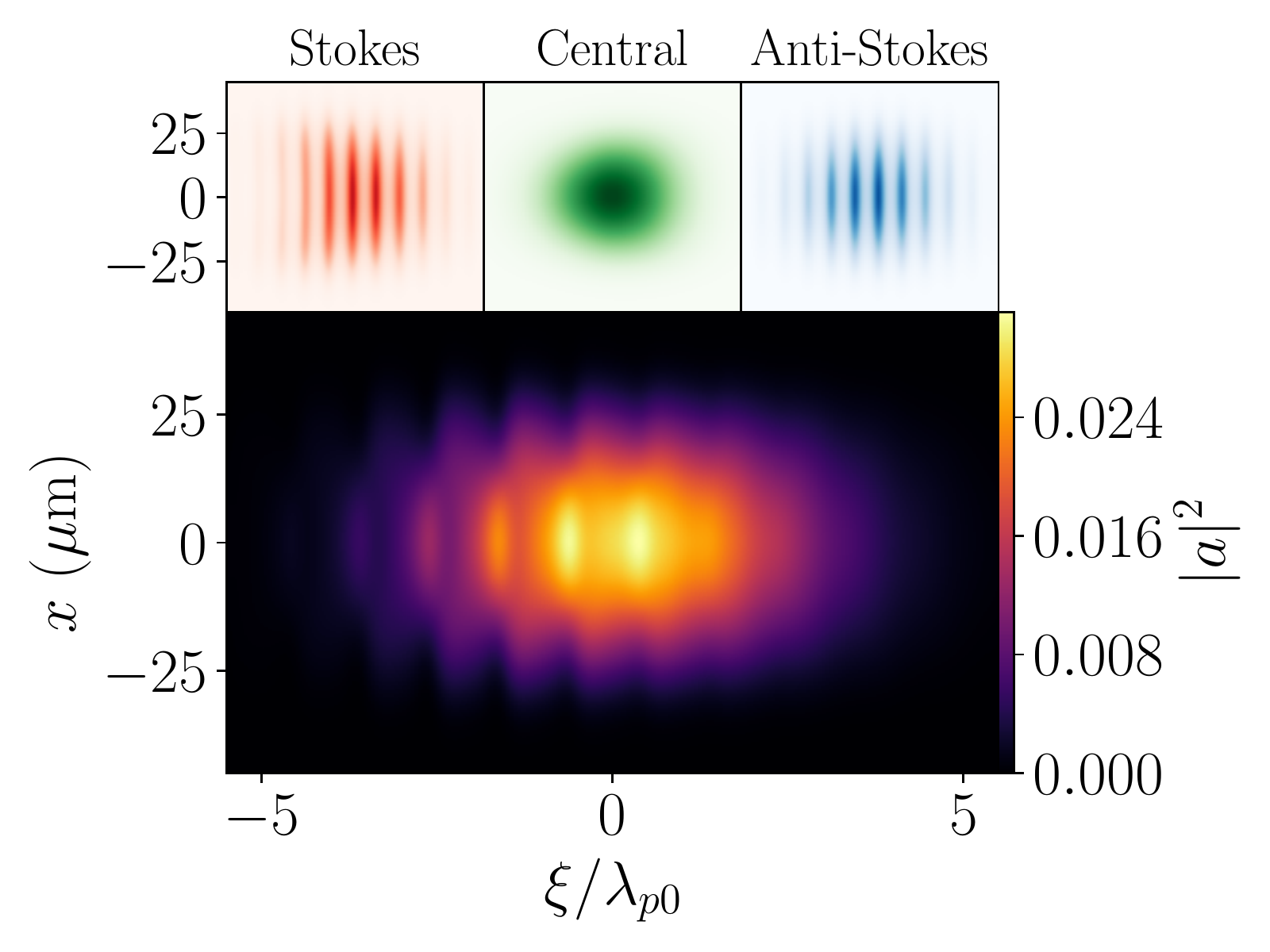}
        }\hspace*{-1.0em}
        \subfloat[\label{subfig:b} $\tau_\text{drive}=\SI{1}{ps}$,\, $W_\text{drive}=\SI{1.2}{J}$]{
            \includegraphics[width=0.333\linewidth]{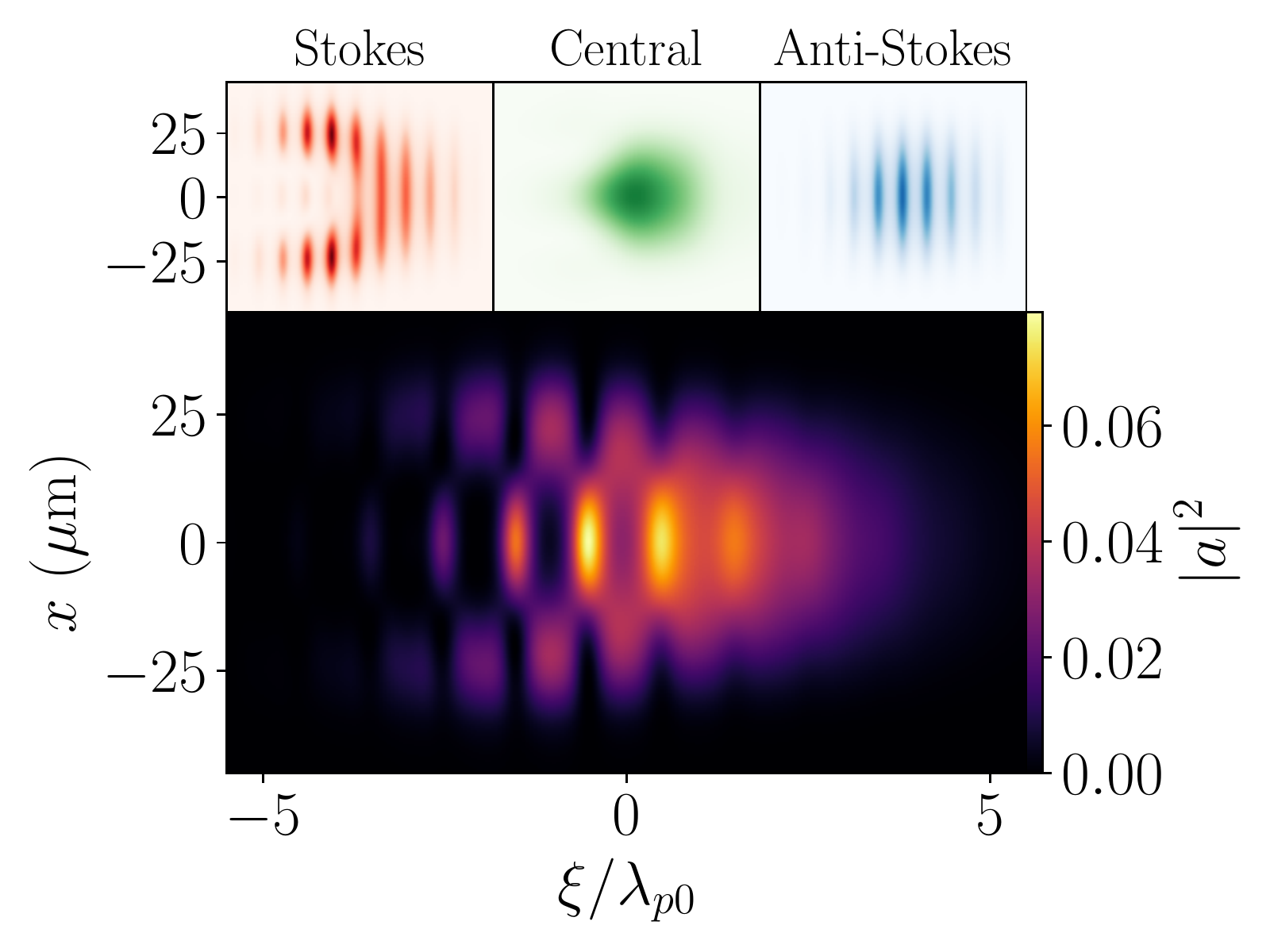}
        }\hspace*{-1.0em}
        \subfloat[\label{subfig:c} $\tau_\text{drive}=\SI{4}{ps}$,\, $W_\text{drive}=\SI{2.4}{J}$]{
            \includegraphics[width=0.333\linewidth]{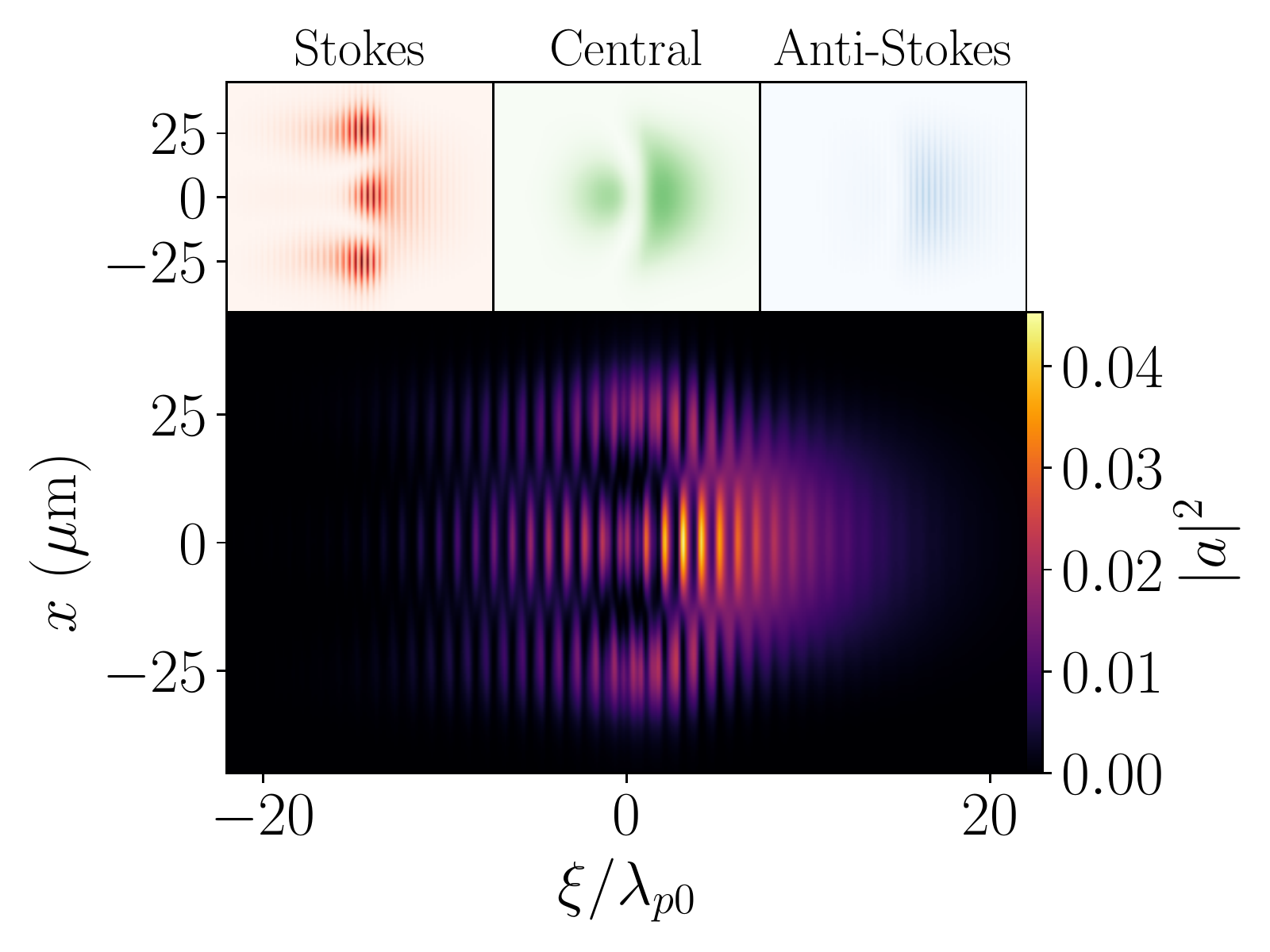}
        }%\hfill
    \caption[Defocusing of Stokes sidebands]{[Color online]. 2D PIC simulations of the intensity profiles, $|a|^2$, of drive pulses at the exit of the modulator with a channel of square cross-section with $w_0 = \SI{30}{\micro m}$ and $W_\text{seed}=\SI{50}{mJ}$, for various energies and durations of the drive pulse. The top row displays the intensity profiles decomposed into its redshifted Stokes $(\omega-\omega_L)<-\omega_{p0}/2$, central $|\omega-\omega_L|<\omega_{p0}/2$, and blueshifted anti-Stokes $(\omega-\omega_L)>\omega_{p0}/2$ components; the bottom panel displays the full intensity profile of the drive pulse.}
    \label{fig:1200mJ}
\end{figure*}

Increasing the drive pulse energy $W_\text{drive}$ eventually disrupts the plasma modulator with a TMI via a self-modulation mechanism. This instability drives time-varying wakes which excite higher order transverse channel modes. The results of PIC simulations demonstrating this phenomenon are given in Fig.\ \ref{fig:1200mJ}. It can be seen that the Stokes sidebands become more defocused towards the trailing end of the pulse, thereby exciting higher order transverse modes, whereas the anti-Stokes-shifted light remains relatively well focused. This asymmetry between the Stokes and anti-Stokes light leads to the formation of low-contrast pulse trains in the modulator, with the Stokes-shifted radiation forming a train off-axis, and the anti-Stokes light forming a train on-axis. This effect becomes worse with increasing drive pulse energy, and can be seen to be especially bad for the \SI{2.4}{J} pulse, which has undergone severe transverse break-up and has been strongly redshifted. This difference in behaviour between the Stokes and anti-Stokes sidebands has previously been  observed as the result of relativistic effects \cite{doi:10.1063/1.859401}. As shown in the Supplemental material \cite{supp}, a similar effect is predicted in the non-relativistic regime when non-paraxial effects are accounted for.

\subsection{Envelope Self-Modulation and Raman Forward Scattering}

We now consider the stability of picosecond-scale pulses propagating in long \emph{unperturbed} plasma channels, i.e.\ in the absence of a wake driven by a seed pulse. As discussed by Mori \cite{641309}, as long as $1/k_Lw_0\gg\omega_p^2/\omega_L^2$ is satisfied, the following parameter determines whether a laser pulse will be dominated by Raman forward scattering (RFS) or envelope self-modulation (SM) instabilities:
\begin{align}
    \Gamma \equiv \frac{P}{1 \text{ TW}}\cdot\frac{\tau_L}{1 \text{ ps}}\cdot\left(\frac{n_e}{1\times10^{19}\text{ cm}^{-3}}\right)^{5/2}\cdot\left(\frac{\lambda_L}{1 \text{ $\mu$m}}\right)^4
\end{align}
where $P$ is the peak laser power and $\tau_L$ is the laser duration. When $\Gamma\geq3$, RFS dominates, whereas SM dominates when $\Gamma\leq0.4$. Substituting laser-plasma parameters used by Jakobsson et al \cite{PhysRevLett.127.184801}, we find $\Gamma = 6.3\times10^{-5}$ and $1/k_Lw_0=55\times10^{-4}\gg\omega_p^2/\omega_L^2=2.4\times10^{-4}$, meaning that we are well in the SM-dominated regime. The maximum growth rate and $e$-folding number of envelope SM in a uniform plasma is given by \cite{PhysRevLett.33.209,641305}
\begin{align}
    &\gamma_\text{SM} = \frac{1}{8}a_0^2\frac{\omega_p^2}{\omega_L}(1+a_0^2)^{-3/2},\,\,\,\, N_{e,\text{SM}} = \gamma_\text{SM}T_\text{int}
\end{align}
where $T_\text{int} = L_\text{mod}/c$ is the interaction time.The parameters used by Jakobsson et al \cite{PhysRevLett.127.184801} yield $\sim1.3$ $e$-foldings, and hence SM could become problematic, especially if we try to include more energy in the drive pulse.

\subsection{Transverse Mode Stability Condition}

For a pulse propagating in a waveguide, the growth of SM is complicated by oscillations in the spot size of the mode, which arise from excitation of more than one waveguide mode. This becomes relevant when the SM growth rate is slower than the spot size oscillation frequency \cite{doi:10.1063/1.3357175} $\gamma_\text{SM}<\omega_w=4c^2/\omega_Lw_0^2$. In this section we consider the effects of the plasma channel on self-modulation.

Consider coupling a slightly unmatched drive pulse into a parabolic plasma channel so it undergoes small spot size oscillations, and assume that no centroid oscillations are present, so that only radial Laguerre-Gaussian modes $\text{LG}_{p0}(r)$ are excited (see Supplemental Material \cite{supp}). Neglecting relativistic effects, applying time-dependent perturbation theory to Eq. (\ref{eq:GNLS}), the coefficients $\alpha_p(\xi,\tau)$ of each radial Laguerre-Gaussian mode $p$ at longitudinal coordinate $\xi$ are found to evolve according to:
\begin{align}\label{eq:TDPT}
    &\frac{i}{\omega_c}\frac{\partial \alpha_p(\xi,\tau)}{\partial\tau} = \nonumber \\
    &\sum_n \alpha_n(\xi,\tau)\Big\langle\text{LG}_{p0}\Big|\frac{\delta n+\delta n_\text{NL}}{\Delta n}\Big|\text{LG}_{n0}\Big\rangle e^{i(p-n)\omega_w\tau} \,, \nonumber \\
    &a(r,\xi,\tau) = e^{-i\omega_c\tau}\sum_p \alpha_p(\xi,\tau)\text{LG}_{p0}(r)e^{-ip\omega_w\tau}
\end{align}
where $\delta n$ is the fixed seed wake, $\delta n_\text{NL}$ is the self-wake of the drive pulse, and $\omega_w$ and $\omega_c=\omega_w/2$ are the spot size and centroid oscillation frequencies respectively \cite{doi:10.1063/1.3357175}. Coupling the drive pulse into a slightly unmatched channel corresponds to following set of initial conditions
\begin{align}
    \alpha_0 = a_0f(\xi),\quad\alpha_1 = \epsilon_wa_0f(\xi),\quad\alpha_{p\neq0,1} \approx 0
\end{align}
where $\epsilon_w=-\delta w/w_0\ll1$ is the channel spot size mismatch parameter. In order to solve Eq. (\ref{eq:TDPT}), the self-wake $\delta n_\text{NL}$ must be known. We estimate the self-wake as follows. We first assume that the self-wake can be neglected, and calculate the intensity modulation of the drive pulse caused by the seed wake only. We then use this intensity modulation to calculate the self-wake it would excite. This estimate of the self-wake can then be used to define the plasma modulator stability condition, which sets bounds on the laser-plasma parameters to prevent nonlinear self-modulation from exciting transverse mode transitions.

We will work in the shallow channel limit $\Delta n\ll n_{00}$, which allows us to neglect the effects of the channel on wake structure \cite{doi:10.1063/1.872186}. This gives a seed wake of the form \cite{doi:10.1063/1.860707}
\begin{align}\label{eq:seedwake}
    \delta n(r,\xi) = \delta n_s\cos(k_{p0}\xi)\text{LG}_{00}^2(r)
\end{align}
where $k_{p0}=\omega_{p0}/v_g$ and $\delta n_s$ denotes the on-axis seed wake amplitude. As the drive pulse should remain primarily in the fundamental mode, we can approximate Eq. (\ref{eq:TDPT}) as a two-level system comprising the $\text{LG}_{00}$ and $\text{LG}_{10}$ modes with transitions between them driven by the seed wake
\begin{align}\label{eq:2level}
    i\frac{\partial \alpha_0}{\partial\tau} &= 2\Omega_s\cos(k_{p0}\xi)\left(\alpha_0+\tfrac{1}{2}\alpha_1e^{-i\omega_w\tau}\right)\,, \nonumber \\
    i\frac{\partial \alpha_1}{\partial\tau} &= 2\Omega_s\cos(k_{p0}\xi)\left(\tfrac{1}{2}\alpha_0e^{i\omega_w\tau}+\tfrac{1}{2}\alpha_1\right)
\end{align}
where $\Omega_s=(\omega_{p0}^2/8\omega_L)(\delta n_s/n_{00})$ is the rate of spectral modulation parameter. Note that since we are using the paraxial description, we are implicitly assuming symmetry between the Stokes and anti-Stokes sidebands (see Supplemental Material \cite{supp} for the non-paraxial description). We can already see from these expressions that the first radial mode spectrally modulates half as fast as the fundamental, as it is more sensitive to the radial drop-off of the seed wake amplitude $\sim\text{LG}_{00}^2(r)$. This asymmetry, coupled to spot size oscillations, is one of two effects contributing to plasma-resonant modulations of the drive pulse intensity which excite a self-wake. Since the spot size oscillation frequency is necessarily much higher than the spectral modulation rate, i.e.\ $\omega_w\gg\Omega_s$, as a consequence of the seed wake being small relative to the channel depth parameter, we can integrate Eq. (\ref{eq:2level}) by assuming most of the light remains in the fundamental mode to find a first order solution $\alpha_0 = \alpha_0^{(0)}(\xi,\tau)+\alpha_0^{(1)}(\xi,\tau),\,\alpha_1 = \alpha_1^{(1)}(\xi,\tau)$. This yields the following intensity modulation
\begin{widetext}
\begin{align}\label{eq:intensity_mod}
    \frac{|a(r,\xi,\tau)|^2}{|a_0|^2f^2(\xi)} = \left\{1-2\left(\epsilon_w+\frac{\Omega_s\cos(k_{p0}\xi)}{\omega_w}\right)\left(\cos[\Omega_s\tau\cos(k_{p0}\xi)]-1\right)\right\}&\text{LG}_{00}^2(r) \nonumber \\
    + \left\{2\epsilon_w\cos\left[\Omega_s\tau\cos(k_{p0}\xi)-\omega_w\tau\right]
    + \frac{2\Omega_s\cos(k_{p0}\xi)}{\omega_w}\left(\cos\left[\Omega_s\tau\cos(k_{p0}\xi)-\omega_w\tau\right]-1\right)\right\}&\text{LG}_{00}(r)\text{LG}_{10}(r)\,.
\end{align}
\end{widetext}
We see from Eq. (\ref{eq:intensity_mod}) that the drive pulse is modulated radially, longitudinally, and temporally. This modulation can be physically understood by splitting it into three effects which can be isolated by setting certain terms to zero. The first is a longitudinally uniform spot size oscillation of amplitude $\delta w$, which can be recovered by setting $\Omega_s=0$. The second comes from coupling between the $\delta w$ spot size oscillations and the seed wake spectral modulation due to the effect mentioned earlier where higher order radial modes spectrally modulate slower than the fundamental mode in the $\sim\text{LG}_{00}^2(r)$ seed wake. The third depends purely on the seed wake, which can be seen when setting $\epsilon_w=0$. The seed wake introduces local variations in the matched spot size as it perturbs the channel plasma density, varying the spot size longitudinally in $\xi$ and temporally in $\tau$.

\begin{figure}[tb]
    \centering
        \vspace*{-1.4em}
        \hspace*{-1.1em}
        \subfloat[\label{subfig:250fs} $\tau_\text{drive}=\SI{250}{fs}$]{
            \includegraphics[width=0.52\linewidth]{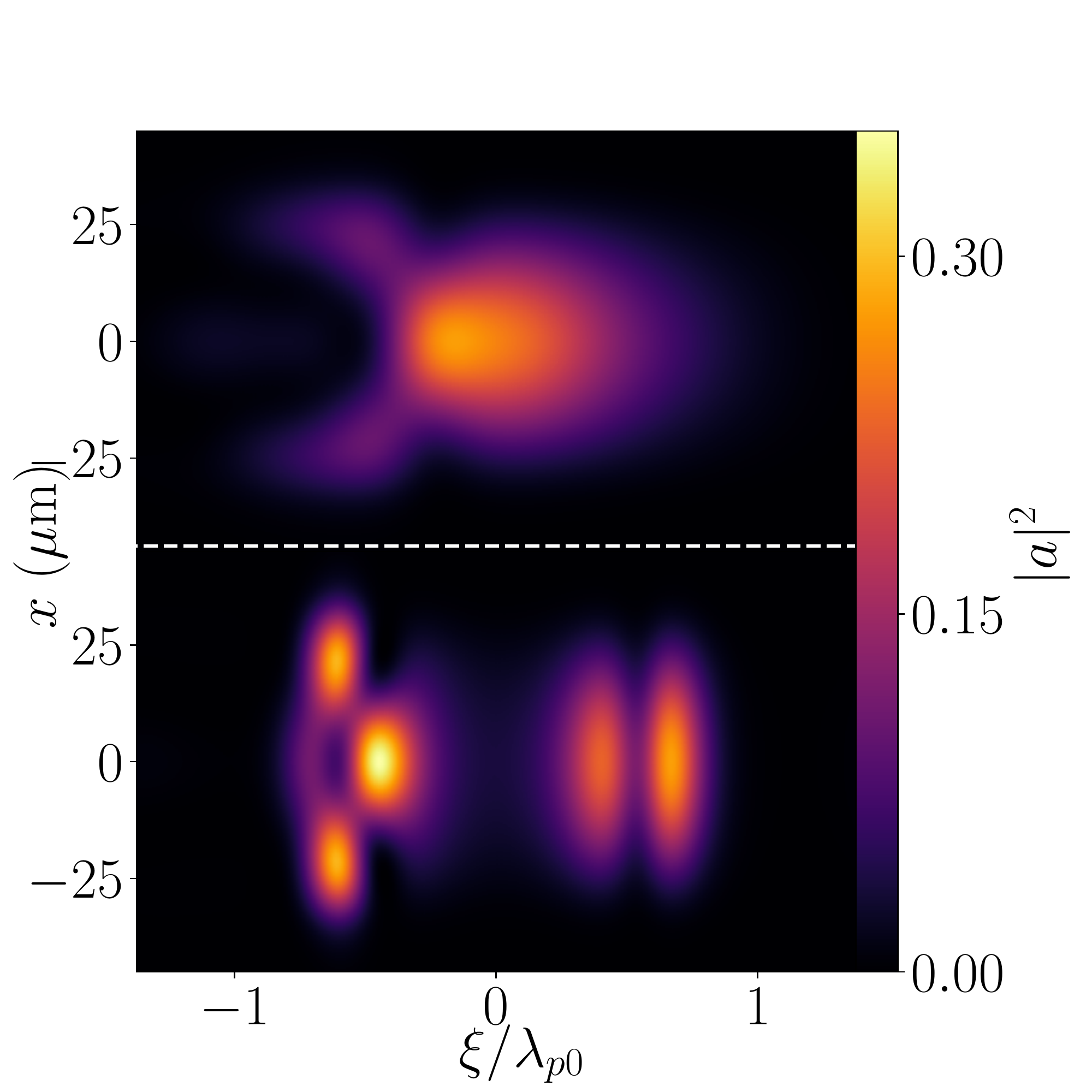}
        }\hspace*{-1.1em}
        \subfloat[\label{subfig:d} $\tau_\text{drive}=\SI{1}{ps}$]{
            \includegraphics[width=0.52\linewidth]{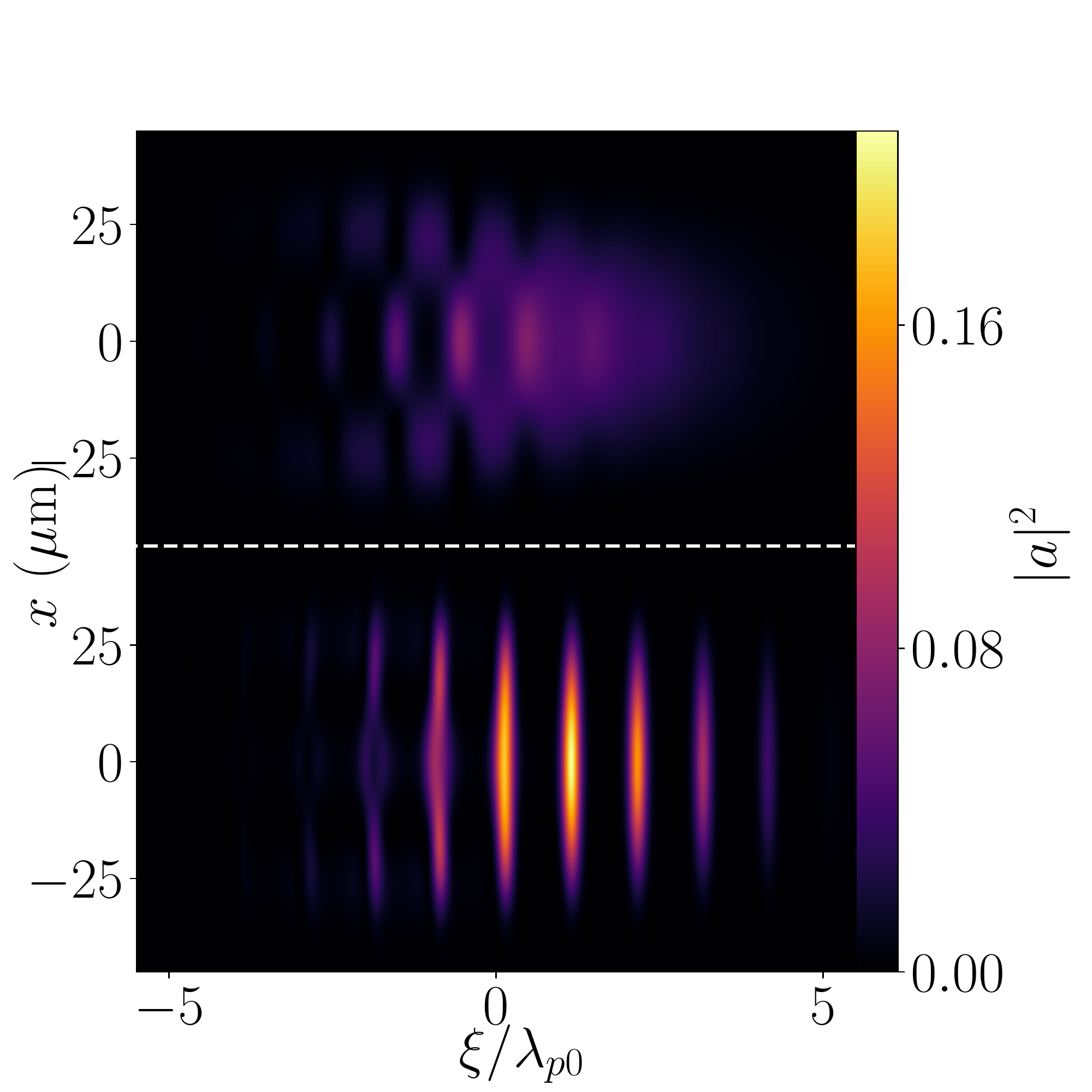}
        }\hfill
        \vspace*{-1.4em}
        \hspace*{-1.1em}
        \subfloat[\label{subfig:e} $\tau_\text{drive}=\SI{2}{ps}$]{
            \includegraphics[width=0.52\linewidth]{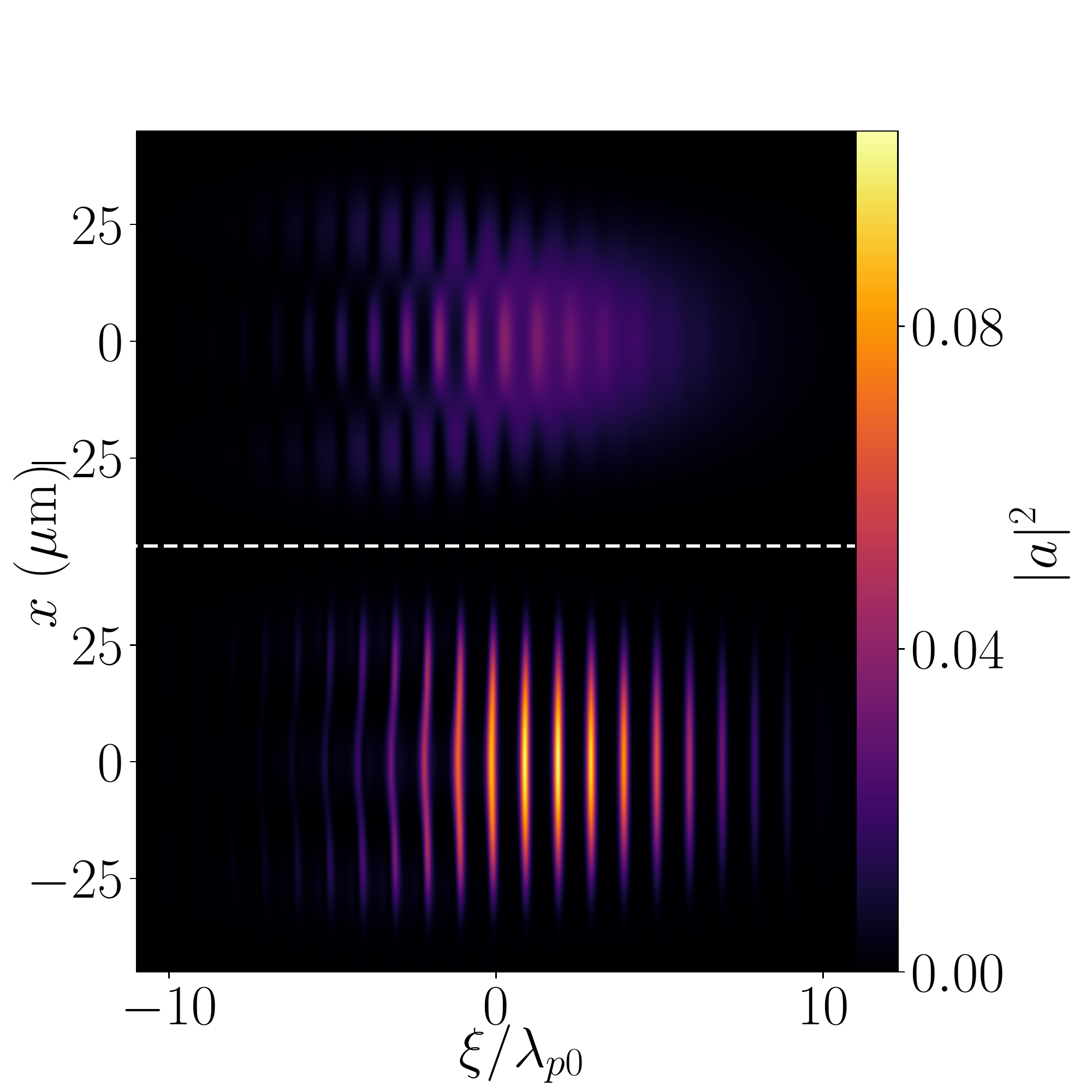}
        }\hspace*{-1.1em}
        \subfloat[\label{subfig:f} $\tau_\text{drive}=\SI{4}{ps}$]{
            \includegraphics[width=0.52\linewidth]{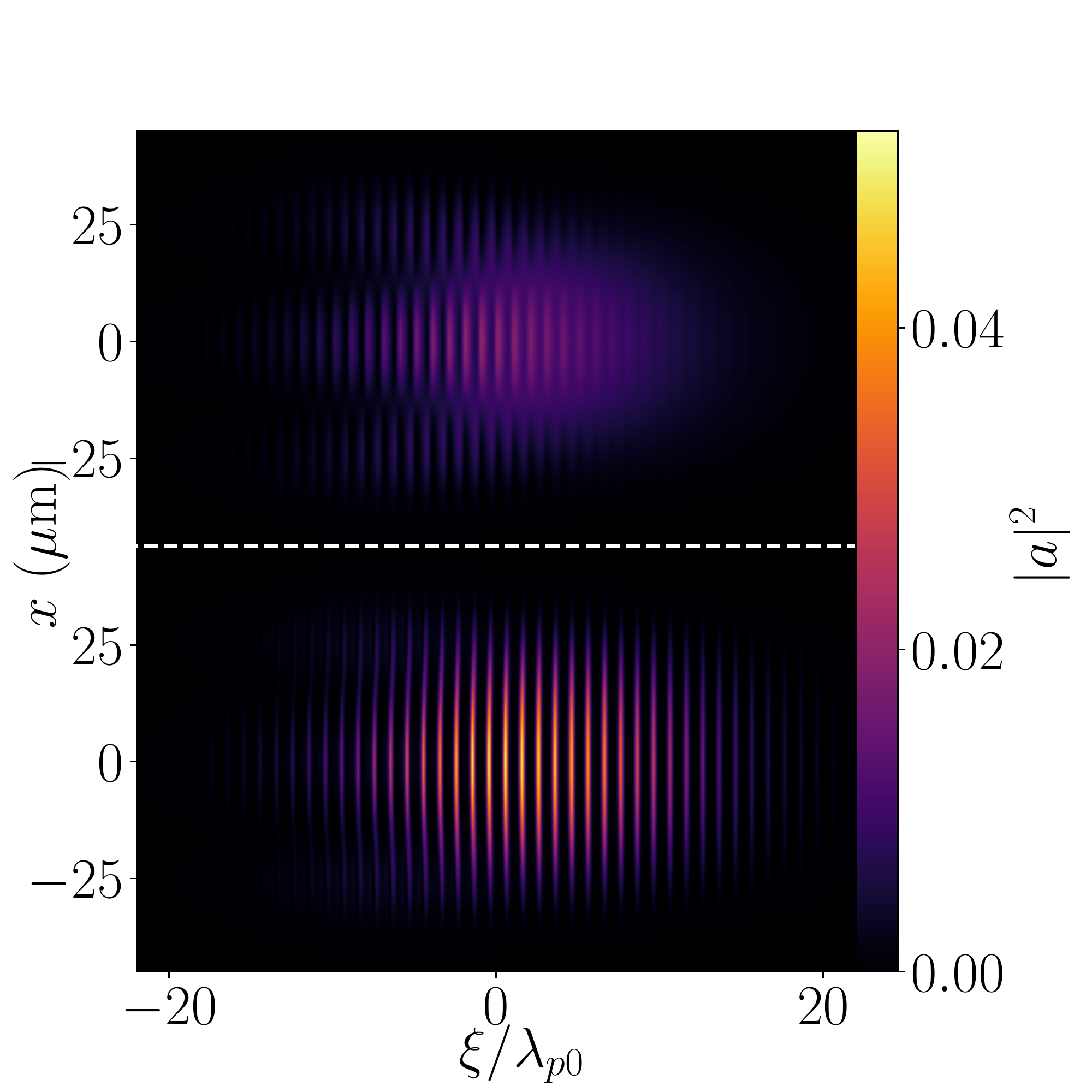}
        }%\hfill
    \caption{Calculated intensity profiles, $|a|^2$, of drive pulses at the exit of the modulator before (top) and after (bottom) compression into a pulse train for drive pulses of energy $W_\text{drive}=\SI{1.2}{J}$ and FWHM duration: (a) \SI{0.25}{ps}, (b) \SI{1}{ps}, (c) \SI{2}{ps} and (d) \SI{4}{ps}. For these 2D PIC simulations the modulator was taken to have a square cross-section with $w_0 = \SI{30}{\micro m}$, and the seed pulse to have an energy of $W_\text{seed}=\SI{50}{mJ}$. }
    \label{fig:selfmod}
\end{figure}

To estimate the self-wake $\delta n_\text{NL}$ that would be excited by this intensity modulation, we are only interested in keeping terms resonant with the plasma $\sim\cos(k_{p0}\xi+\phi)$ which excite the largest amplitude self-wake. We also only consider propagation times up to $\Omega_s\tau_\text{mod}\sim1/2$, as this provides sufficient spectral modulation for compression into a pulse train which roughly coincides with the minimum modulation required to reach the accelerator stage wake amplitude plateau discussed by Jakobsson et al \cite{PhysRevLett.127.184801}. With both of these considerations in mind, we can Taylor expand the $\cos\left[\Omega_s\tau\cos(k_{p0}\xi)-\omega_w\tau\right]$ terms in Eq. (\ref{eq:intensity_mod}) to first order in $\Omega_s\tau$. This gives the plasma-resonant part of the intensity modulation and the approximate self-wake it would excite \cite{PhysRevLett.29.701}
\begin{widetext}
\begin{align}\label{eq:dn_NL}
    |a(r,\xi,\tau)|_\text{res}^2 = |a_0|^2f^2(\xi)&\left(2\epsilon_w\Omega_s\tau\sin(\omega_w\tau) + \frac{2\Omega_s\left[\cos(\omega_w\tau)-1\right]}{\omega_w}\right)\cos(k_{p0}\xi)\text{LG}_{00}(r)\text{LG}_{10}(r)\,, \nonumber \\
    \frac{\delta n_\text{NL}(r,\xi,\tau)}{n_{00}} = 
    \frac{e^2}{8\pi^2m_e^2\epsilon_0c^5}\frac{\omega_{p0}W_\text{drive}(\xi)\lambda_L^2}{\pi w_0^2} &\left(2\epsilon_w\Omega_s\tau\sin(\omega_w\tau) + \frac{2\Omega_s\left[\cos(\omega_w\tau)-1\right]}{\omega_w}\right)\sin(k_{p0}\xi)\text{LG}_{00}(r)\text{LG}_{10}(r)
\end{align}
\end{widetext}
where $W_\text{drive}(\xi)$ indicates the total energy of the drive pulse contained between its head and coordinate $\xi$. To prevent self-modulation driving transverse mode transitions (and to make this calculation self-consistent), we require that the self-wake effect on the $\text{LG}_{00}\rightarrow\text{LG}_{10}$ transition must be negligible throughout the full propagation in the modulator, resulting in the constraint
\begin{align}
    \int_0^{\tau_\text{mod}}\!\!\!\frac{d\tau}{\tau_\text{mod}}\big\langle\text{LG}_{10}\big|\delta n_\text{NL}e^{i\omega_w\tau}\big|\text{LG}_{00}\big\rangle \ll \big\langle\text{LG}_{10}\big|\delta n\big|\text{LG}_{00}\big\rangle\,.
\end{align}
Substituting $\Omega_s\tau_\text{mod}=1/2$ and Eq. (\ref{eq:dn_NL}) into this constraint yields the plasma modulator transverse mode stability condition
\begin{align}\label{eq:amod_analytic}
    &\left|\frac{\delta w/w_0}{\delta n_s/n_{00}} + \frac{k_{p0}^2w_0^2}{8}\right|\frac{\omega_{p0}W_\text{drive}\lambda_L^2}{\pi w_0^2} \ll P_\text{mod}\,, \nonumber \\
    &P_\text{mod} = \frac{32\pi^2m_e^2\epsilon_0c^5}{e^2} \approx \SI{220}{GW}\,.
\end{align}
This sets a limit on the total energy of the drive pulse $W_\text{drive}$ for a given plasma density, spot size, laser wavelength, seed wake and channel spot size mismatch. For the laser-plasma parameters used by Jakobsson et al \cite{PhysRevLett.127.184801}, taking  $\delta n_s/n_{00}\sim2.5\%$ and $\delta w/w_0\sim 12\%$ from the PIC simulations therein, the requirement of Eq. (\ref{eq:amod_analytic}) becomes $\SI{37}{GW} \ll \SI{220}{GW}$, which is satisfied. Hence, in this regime we do not expect the self-modulation effects to be debilitating to the plasma modulator. In PIC simulations, we have found that even letting the LHS of Eq. (\ref{eq:amod_analytic}) go up to $\sim\,$\SI{70}{GW} remains stable enough for compression into a pulse train despite the excitation of the first radial mode being non-negligible. Figure \ref{fig:selfmod} shows the calculated intensity profiles, $|a|^2$, of drive pulses at the exit of the modulator before and after compression into a pulse train (by removing the expected \cite{PhysRevLett.127.184801_supp} sideband spectral phase $\psi_m$ of a pulse without TMI (see Supplemental Material \cite{supp})) for drive pulses of energy $W_\text{drive}=\SI{1.2}{J}$ and various pulse durations. It can be seen that, despite the pulse duration varying by a factor of 16, they each undergo the same amount of spot size oscillation driven transitions to higher order transverse modes. This is in agreement with Eq. (\ref{eq:amod_analytic}), which is independent of the drive pulse duration. As a consequence, for drive pulses of duration 1, 2 and \SI{4}{ps}, the transverse structure of the pulse at the end of the modulator is identical. For drive pulses of \SI{250}{fs} duration, the structure is also similar, but in this case the pulse duration is only approximately two plasma periods long, and hence the assumption of small bandwidth breaks down. We note also that Fig.\ \ref{fig:selfmod} shows that in each case the spectrally-modulated drive pulse can still be compressed into a well-defined pulse train suitable for the accelerator stage, despite the transverse structure that it has developed.

For most practical applications, the spot size oscillations will be determined by the mismatch between the transverse amplitude profile of the lowest-order channel mode, and that of the drive pulse at the channel entrance. However, even in the limit of a perfectly matched channel, small spot size oscillations can arise from other sources, such as ponderomotive and relativistic self-focusing \cite{SIEGRIST1976402,doi:10.1063/1.860828}. In addition, the $k_{p0}^2w_0^2/8$ term in Eq. (\ref{eq:amod_analytic}), which comes from the seed wake forming a plasma-resonant variation in matched spot size, ensures that there will always be an upper limit on the drive pulse energy.

There are two ways that excitation of  higher order transverse modes can be mitigated, other than ensuring that Eq. (\ref{eq:amod_analytic}) is satisfied. However, each of these comes at a cost. First, the treatment above assumed propagation in a shallow channel. For deeper channels, i.e.\ $\Delta n\sim n_{00}$, the self-wake will be partially suppressed by the off-resonant plasma wave, and the difference in spectral modulation rate of the higher-order mode and the fundamental will be decreased due to the radial component of the seed wake. This allows for more energy to propagate in the modulator without transitions to higher order modes. However, channels of this form also suppress the spectral modulation towards the tail of the pulse, as shown in Fig.\ \ref{fig:specmodsupp2}, which is detrimental to pulse compression. Another option would be to use a leaky channel that leaks higher order modes faster than the fundamental mode \cite{doi:10.1063/1.4975860,doi:10.1063/1.5006198}. However, this approach would have reduced efficiency, since drive pulse energy transferred to higher order modes would be lost.

\section{Conclusion}

We have derived a full 3D analytic theory of seeded spectral modulation, and have used this to establish the useful operating regime for the modulator stage in the P-MoPA. This model is found to be in very good agreement with those obtained from 2D PIC simulations of the modulator.

The analytic theory leads to several important conclusions. First, the spectral modulation of the drive pulse is independent of radial distance from the axis of the modulator channel. This ensures that, after leaving the modulator, the entire spectrally-modulated pulse can be compressed by a simple optical system that removes the spectral phase accumulated in the modulator. Second, curvature of the seed-pulse-driven plasma waves is shown to reduce the degree of spectral modulation, and hence the modulation of the pulse train that is generated after compression. This finding establishes limits on the shape of the channel used in a P-MoPA modulator.

We also explored limits to the operating parameters of a seeded modulator set by the self-modulation of the drive pulse and excitation of higher-order transverse channel modes. We found that the operation of the modulator is limited by the onset of the transverse mode instability (TMI), similar to the TMI observed in high power fiber laser systems. An analysis of the excitation of higher-order modes allowed the identification of a condition on the energy of the drive pulse, the relative amplitude of oscillations in its spot size, and the relative amplitude of the seed-pulse-driven wake, that must be satisfied for stable operation.

Finally we emphasize that the results presented here show that the modulator in a P-MoPA can exhibit stable operation over a much broader range of operating parameters than considered in the original proposal \cite{PhysRevLett.127.184801}. This includes operation at higher drive pulse energies, which bodes well for the development of high-repetition-rate, GeV-scale P-MoPAs.

\acknowledgements

This work was supported by the UK Engineering and Physical Sciences Research Council (EPSRC) (Grant No.\ EP/V006797/1), the UK Science and Technologies Facilities Council (Grant No.\ ST/V001655/1]), InnovateUK (Grant No.\ 10059294), UKRI funding (ARCHER2 Pioneer Projects)  and the Ken and Veronica Tregidgo Scholarship in Atomic and Laser Physics. This publication arises from research funded by the John Fell Oxford University Press Research Fund. This work used the ARCHER2 UK National Supercomputing Service (ARCHER2 PR17125)  \url{https://www.archer2.ac.uk}. This research used the open-source particle-in-cell code WarpX \cite{WarpX} \url{https://github.com/ECP-WarpX/WarpX}, primarily funded by the US DOE Exascale Computing Project. Primary WarpX contributors are with LBNL, LLNL, CEA-LIDYL, SLAC, DESY, CERN, and Modern Electron. We acknowledge all WarpX contributors.

This research was funded in whole, or in part, by EPSRC and STFC, which are Plan S funders. For the purpose of Open Access, the author has applied a CC BY public copyright licence to any Author Accepted Manuscript version arising from this submission.

The input decks used for the PIC simulations presented in this paper are available at \url{http://doi.org/10.5281/zenodo.7734261}

\bibliography{references}% Produces the bibliography via BibTeX.

%apsrev4-2.bst 2019-01-14 (MD) hand-edited version of apsrev4-1.bst
%Control: key (0)
%Control: author (8) initials jnrlst
%Control: editor formatted (1) identically to author
%Control: production of article title (0) allowed
%Control: page (0) single
%Control: year (1) truncated
%Control: production of eprint (0) enabled
\providecommand{\noopsort}[1]{}\providecommand{\singleletter}[1]{#1}%
\begin{thebibliography}{39}%
\makeatletter
\providecommand \@ifxundefined [1]{%
 \@ifx{#1\undefined}
}%
\providecommand \@ifnum [1]{%
 \ifnum #1\expandafter \@firstoftwo
 \else \expandafter \@secondoftwo
 \fi
}%
\providecommand \@ifx [1]{%
 \ifx #1\expandafter \@firstoftwo
 \else \expandafter \@secondoftwo
 \fi
}%
\providecommand \natexlab [1]{#1}%
\providecommand \enquote  [1]{``#1''}%
\providecommand \bibnamefont  [1]{#1}%
\providecommand \bibfnamefont [1]{#1}%
\providecommand \citenamefont [1]{#1}%
\providecommand \href@noop [0]{\@secondoftwo}%
\providecommand \href [0]{\begingroup \@sanitize@url \@href}%
\providecommand \@href[1]{\@@startlink{#1}\@@href}%
\providecommand \@@href[1]{\endgroup#1\@@endlink}%
\providecommand \@sanitize@url [0]{\catcode `\\12\catcode `\$12\catcode
  `\&12\catcode `\#12\catcode `\^12\catcode `\_12\catcode `\%12\relax}%
\providecommand \@@startlink[1]{}%
\providecommand \@@endlink[0]{}%
\providecommand \url  [0]{\begingroup\@sanitize@url \@url }%
\providecommand \@url [1]{\endgroup\@href {#1}{\urlprefix }}%
\providecommand \urlprefix  [0]{URL }%
\providecommand \Eprint [0]{\href }%
\providecommand \doibase [0]{https://doi.org/}%
\providecommand \selectlanguage [0]{\@gobble}%
\providecommand \bibinfo  [0]{\@secondoftwo}%
\providecommand \bibfield  [0]{\@secondoftwo}%
\providecommand \translation [1]{[#1]}%
\providecommand \BibitemOpen [0]{}%
\providecommand \bibitemStop [0]{}%
\providecommand \bibitemNoStop [0]{.\EOS\space}%
\providecommand \EOS [0]{\spacefactor3000\relax}%
\providecommand \BibitemShut  [1]{\csname bibitem#1\endcsname}%
\let\auto@bib@innerbib\@empty
%</preamble>
\bibitem [{\citenamefont {Tajima}\ and\ \citenamefont
  {Dawson}(1979)}]{PhysRevLett.43.267}%
  \BibitemOpen
  \bibfield  {author} {\bibinfo {author} {\bibfnamefont {T.}~\bibnamefont
  {Tajima}}\ and\ \bibinfo {author} {\bibfnamefont {J.~M.}\ \bibnamefont
  {Dawson}},\ }\bibfield  {title} {\bibinfo {title} {Laser electron
  accelerator},\ }\href {https://doi.org/10.1103/PhysRevLett.43.267} {\bibfield
   {journal} {\bibinfo  {journal} {Phys. Rev. Lett.}\ }\textbf {\bibinfo
  {volume} {43}},\ \bibinfo {pages} {267} (\bibinfo {year} {1979})}\BibitemShut
  {NoStop}%
\bibitem [{\citenamefont {Strickland}\ and\ \citenamefont
  {Mourou}(1985)}]{STRICKLAND1985219}%
  \BibitemOpen
  \bibfield  {author} {\bibinfo {author} {\bibfnamefont {D.}~\bibnamefont
  {Strickland}}\ and\ \bibinfo {author} {\bibfnamefont {G.}~\bibnamefont
  {Mourou}},\ }\bibfield  {title} {\bibinfo {title} {Compression of amplified
  chirped optical pulses},\ }\href
  {https://doi.org/https://doi.org/10.1016/0030-4018(85)90120-8} {\bibfield
  {journal} {\bibinfo  {journal} {Optics Communications}\ }\textbf {\bibinfo
  {volume} {56}},\ \bibinfo {pages} {219} (\bibinfo {year} {1985})}\BibitemShut
  {NoStop}%
\bibitem [{\citenamefont {Dawson}\ \emph {et~al.}(2012)\citenamefont {Dawson},
  \citenamefont {Crane}, \citenamefont {Messerly}, \citenamefont {Prantil},
  \citenamefont {Pax}, \citenamefont {Sridharan}, \citenamefont {Allen},
  \citenamefont {Drachenberg}, \citenamefont {Phan}, \citenamefont {Heebner},
  \citenamefont {Ebbers}, \citenamefont {Beach}, \citenamefont {Hartouni},
  \citenamefont {Siders}, \citenamefont {Spinka}, \citenamefont {Barty},
  \citenamefont {Bayramian}, \citenamefont {Haefner}, \citenamefont {Albert},
  \citenamefont {Lowdermilk}, \citenamefont {Rubenchik},\ and\ \citenamefont
  {Bonanno}}]{doi:10.1063/1.4773687}%
  \BibitemOpen
  \bibfield  {author} {\bibinfo {author} {\bibfnamefont {J.~W.}\ \bibnamefont
  {Dawson}}, \bibinfo {author} {\bibfnamefont {J.~K.}\ \bibnamefont {Crane}},
  \bibinfo {author} {\bibfnamefont {M.~J.}\ \bibnamefont {Messerly}}, \bibinfo
  {author} {\bibfnamefont {M.~A.}\ \bibnamefont {Prantil}}, \bibinfo {author}
  {\bibfnamefont {P.~H.}\ \bibnamefont {Pax}}, \bibinfo {author} {\bibfnamefont
  {A.~K.}\ \bibnamefont {Sridharan}}, \bibinfo {author} {\bibfnamefont {G.~S.}\
  \bibnamefont {Allen}}, \bibinfo {author} {\bibfnamefont {D.~R.}\ \bibnamefont
  {Drachenberg}}, \bibinfo {author} {\bibfnamefont {H.~H.}\ \bibnamefont
  {Phan}}, \bibinfo {author} {\bibfnamefont {J.~E.}\ \bibnamefont {Heebner}},
  \bibinfo {author} {\bibfnamefont {C.~A.}\ \bibnamefont {Ebbers}}, \bibinfo
  {author} {\bibfnamefont {R.~J.}\ \bibnamefont {Beach}}, \bibinfo {author}
  {\bibfnamefont {E.~P.}\ \bibnamefont {Hartouni}}, \bibinfo {author}
  {\bibfnamefont {C.~W.}\ \bibnamefont {Siders}}, \bibinfo {author}
  {\bibfnamefont {T.~M.}\ \bibnamefont {Spinka}}, \bibinfo {author}
  {\bibfnamefont {C.~P.~J.}\ \bibnamefont {Barty}}, \bibinfo {author}
  {\bibfnamefont {A.~J.}\ \bibnamefont {Bayramian}}, \bibinfo {author}
  {\bibfnamefont {L.~C.}\ \bibnamefont {Haefner}}, \bibinfo {author}
  {\bibfnamefont {F.}~\bibnamefont {Albert}}, \bibinfo {author} {\bibfnamefont
  {W.~H.}\ \bibnamefont {Lowdermilk}}, \bibinfo {author} {\bibfnamefont
  {A.~M.}\ \bibnamefont {Rubenchik}},\ and\ \bibinfo {author} {\bibfnamefont
  {R.~E.}\ \bibnamefont {Bonanno}},\ }\bibfield  {title} {\bibinfo {title}
  {High average power lasers for future particle accelerators},\ }\href
  {https://doi.org/10.1063/1.4773687} {\bibfield  {journal} {\bibinfo
  {journal} {AIP Conference Proceedings}\ }\textbf {\bibinfo {volume} {1507}},\
  \bibinfo {pages} {147} (\bibinfo {year} {2012})},\ \Eprint
  {https://arxiv.org/abs/https://aip.scitation.org/doi/pdf/10.1063/1.4773687}
  {https://aip.scitation.org/doi/pdf/10.1063/1.4773687} \BibitemShut {NoStop}%
\bibitem [{\citenamefont {Hidding}\ \emph {et~al.}(2019)\citenamefont
  {Hidding}, \citenamefont {Hooker}, \citenamefont {Jamison}, \citenamefont
  {Muratori}, \citenamefont {Murphy}, \citenamefont {Najmudin}, \citenamefont
  {Pattathil}, \citenamefont {Sarri}, \citenamefont {Streeter}, \citenamefont
  {Welsch}, \citenamefont {Wing},\ and\ \citenamefont {Xia}}]{PWASC}%
  \BibitemOpen
  \bibfield  {author} {\bibinfo {author} {\bibfnamefont {B.}~\bibnamefont
  {Hidding}}, \bibinfo {author} {\bibfnamefont {S.}~\bibnamefont {Hooker}},
  \bibinfo {author} {\bibfnamefont {S.}~\bibnamefont {Jamison}}, \bibinfo
  {author} {\bibfnamefont {B.}~\bibnamefont {Muratori}}, \bibinfo {author}
  {\bibfnamefont {C.}~\bibnamefont {Murphy}}, \bibinfo {author} {\bibfnamefont
  {Z.}~\bibnamefont {Najmudin}}, \bibinfo {author} {\bibfnamefont
  {R.}~\bibnamefont {Pattathil}}, \bibinfo {author} {\bibfnamefont
  {G.}~\bibnamefont {Sarri}}, \bibinfo {author} {\bibfnamefont
  {M.}~\bibnamefont {Streeter}}, \bibinfo {author} {\bibfnamefont
  {C.}~\bibnamefont {Welsch}}, \bibinfo {author} {\bibfnamefont
  {M.}~\bibnamefont {Wing}},\ and\ \bibinfo {author} {\bibfnamefont
  {G.}~\bibnamefont {Xia}},\ }\bibfield  {title} {\bibinfo {title} {Plasma
  wakefield accelerator research 2019–2040: A community-driven uk roadmap
  compiled by the plasma wakefield accelerator steering committee (pwasc)},\
  }\href {https://www.pwasc.org.uk/uk-roadmap} {\bibfield  {journal} {\bibinfo
  {journal} {PWASC}\ } (\bibinfo {year} {2019})}\BibitemShut {NoStop}%
\bibitem [{\citenamefont {Herkommer}\ \emph {et~al.}(2020)\citenamefont
  {Herkommer}, \citenamefont {Kr\"{o}tz}, \citenamefont {Jung}, \citenamefont
  {Klingebiel}, \citenamefont {Wandt}, \citenamefont {Bessing}, \citenamefont
  {Walch}, \citenamefont {Produit}, \citenamefont {Michel}, \citenamefont
  {Bauer}, \citenamefont {Kienberger},\ and\ \citenamefont
  {Metzger}}]{Herkommer:20}%
  \BibitemOpen
  \bibfield  {author} {\bibinfo {author} {\bibfnamefont {C.}~\bibnamefont
  {Herkommer}}, \bibinfo {author} {\bibfnamefont {P.}~\bibnamefont
  {Kr\"{o}tz}}, \bibinfo {author} {\bibfnamefont {R.}~\bibnamefont {Jung}},
  \bibinfo {author} {\bibfnamefont {S.}~\bibnamefont {Klingebiel}}, \bibinfo
  {author} {\bibfnamefont {C.}~\bibnamefont {Wandt}}, \bibinfo {author}
  {\bibfnamefont {R.}~\bibnamefont {Bessing}}, \bibinfo {author} {\bibfnamefont
  {P.}~\bibnamefont {Walch}}, \bibinfo {author} {\bibfnamefont
  {T.}~\bibnamefont {Produit}}, \bibinfo {author} {\bibfnamefont
  {K.}~\bibnamefont {Michel}}, \bibinfo {author} {\bibfnamefont
  {D.}~\bibnamefont {Bauer}}, \bibinfo {author} {\bibfnamefont
  {R.}~\bibnamefont {Kienberger}},\ and\ \bibinfo {author} {\bibfnamefont
  {T.}~\bibnamefont {Metzger}},\ }\bibfield  {title} {\bibinfo {title}
  {Ultrafast thin-disk multipass amplifier with 720 {mJ} operating at kilohertz
  repetition rate for applications in atmospheric research},\ }\href
  {https://doi.org/10.1364/OE.404185} {\bibfield  {journal} {\bibinfo
  {journal} {Opt. Express}\ }\textbf {\bibinfo {volume} {28}},\ \bibinfo
  {pages} {30164} (\bibinfo {year} {2020})}\BibitemShut {NoStop}%
\bibitem [{\citenamefont {Nagel}\ \emph {et~al.}(2021)\citenamefont {Nagel},
  \citenamefont {Metzger}, \citenamefont {Bauer}, \citenamefont {Dominik},
  \citenamefont {Gottwald}, \citenamefont {Kuhn}, \citenamefont {Killi},
  \citenamefont {Dekorsy},\ and\ \citenamefont {Schad}}]{Nagel:21}%
  \BibitemOpen
  \bibfield  {author} {\bibinfo {author} {\bibfnamefont {S.}~\bibnamefont
  {Nagel}}, \bibinfo {author} {\bibfnamefont {B.}~\bibnamefont {Metzger}},
  \bibinfo {author} {\bibfnamefont {D.}~\bibnamefont {Bauer}}, \bibinfo
  {author} {\bibfnamefont {J.}~\bibnamefont {Dominik}}, \bibinfo {author}
  {\bibfnamefont {T.}~\bibnamefont {Gottwald}}, \bibinfo {author}
  {\bibfnamefont {V.}~\bibnamefont {Kuhn}}, \bibinfo {author} {\bibfnamefont
  {A.}~\bibnamefont {Killi}}, \bibinfo {author} {\bibfnamefont
  {T.}~\bibnamefont {Dekorsy}},\ and\ \bibinfo {author} {\bibfnamefont {S.-S.}\
  \bibnamefont {Schad}},\ }\bibfield  {title} {\bibinfo {title} {Thin-disk
  laser system operating above 10 {kW} at near fundamental mode beam quality},\
  }\href {https://doi.org/10.1364/OL.416432} {\bibfield  {journal} {\bibinfo
  {journal} {Opt. Lett.}\ }\textbf {\bibinfo {volume} {46}},\ \bibinfo {pages}
  {965} (\bibinfo {year} {2021})}\BibitemShut {NoStop}%
\bibitem [{\citenamefont {{Produit, Thomas}}\ \emph {et~al.}(2021)\citenamefont
  {{Produit, Thomas}}, \citenamefont {{Walch, Pierre}}, \citenamefont
  {{Herkommer, Clemens}}, \citenamefont {{Mostajabi, Amirhossein}},
  \citenamefont {{Moret, Michel}}, \citenamefont {{Andral, Ugo}}, \citenamefont
  {{Sunjerga, Antonio}}, \citenamefont {{Azadifar, Mohammad}}, \citenamefont
  {{Andr\'e, Yves-Bernard}}, \citenamefont {{Mahieu, Beno\^{\i}t}},
  \citenamefont {{Haas, Walter}}, \citenamefont {{Esmiller, Bruno}},
  \citenamefont {{Fournier, Gilles}}, \citenamefont {{Kr\"otz, Peter}},
  \citenamefont {{Metzger, Thomas}}, \citenamefont {{Michel, Knut}},
  \citenamefont {{Mysyrowicz, Andr\'e}}, \citenamefont {{Rubinstein, Marcos}},
  \citenamefont {{Rachidi, Farhad}}, \citenamefont {{Kasparian, J\'er\^ome}},
  \citenamefont {{Wolf, Jean-Pierre}},\ and\ \citenamefont {{Houard,
  Aur\'elien}}}]{Produit:21}%
  \BibitemOpen
  \bibfield  {author} {\bibinfo {author} {\bibnamefont {{Produit, Thomas}}},
  \bibinfo {author} {\bibnamefont {{Walch, Pierre}}}, \bibinfo {author}
  {\bibnamefont {{Herkommer, Clemens}}}, \bibinfo {author} {\bibnamefont
  {{Mostajabi, Amirhossein}}}, \bibinfo {author} {\bibnamefont {{Moret,
  Michel}}}, \bibinfo {author} {\bibnamefont {{Andral, Ugo}}}, \bibinfo
  {author} {\bibnamefont {{Sunjerga, Antonio}}}, \bibinfo {author}
  {\bibnamefont {{Azadifar, Mohammad}}}, \bibinfo {author} {\bibnamefont
  {{Andr\'e, Yves-Bernard}}}, \bibinfo {author} {\bibnamefont {{Mahieu,
  Beno\^{\i}t}}}, \bibinfo {author} {\bibnamefont {{Haas, Walter}}}, \bibinfo
  {author} {\bibnamefont {{Esmiller, Bruno}}}, \bibinfo {author} {\bibnamefont
  {{Fournier, Gilles}}}, \bibinfo {author} {\bibnamefont {{Kr\"otz, Peter}}},
  \bibinfo {author} {\bibnamefont {{Metzger, Thomas}}}, \bibinfo {author}
  {\bibnamefont {{Michel, Knut}}}, \bibinfo {author} {\bibnamefont
  {{Mysyrowicz, Andr\'e}}}, \bibinfo {author} {\bibnamefont {{Rubinstein,
  Marcos}}}, \bibinfo {author} {\bibnamefont {{Rachidi, Farhad}}}, \bibinfo
  {author} {\bibnamefont {{Kasparian, J\'er\^ome}}}, \bibinfo {author}
  {\bibnamefont {{Wolf, Jean-Pierre}}},\ and\ \bibinfo {author} {\bibnamefont
  {{Houard, Aur\'elien}}},\ }\bibfield  {title} {\bibinfo {title} {The laser
  lightning rod project},\ }\href {https://doi.org/10.1051/epjap/2020200243}
  {\bibfield  {journal} {\bibinfo  {journal} {Eur. Phys. J. Appl. Phys.}\
  }\textbf {\bibinfo {volume} {93}},\ \bibinfo {pages} {10504} (\bibinfo {year}
  {2021})}\BibitemShut {NoStop}%
\bibitem [{\citenamefont {Paschotta}\ \emph {et~al.}(2001)\citenamefont
  {Paschotta}, \citenamefont {Aus~der Au}, \citenamefont {Spühler},
  \citenamefont {Erhard}, \citenamefont {Giesen},\ and\ \citenamefont
  {Keller}}]{Paschotta2001}%
  \BibitemOpen
  \bibfield  {author} {\bibinfo {author} {\bibfnamefont {R.}~\bibnamefont
  {Paschotta}}, \bibinfo {author} {\bibfnamefont {J.}~\bibnamefont {Aus~der
  Au}}, \bibinfo {author} {\bibfnamefont {G.}~\bibnamefont {Spühler}},
  \bibinfo {author} {\bibfnamefont {S.}~\bibnamefont {Erhard}}, \bibinfo
  {author} {\bibfnamefont {A.}~\bibnamefont {Giesen}},\ and\ \bibinfo {author}
  {\bibfnamefont {U.}~\bibnamefont {Keller}},\ }\bibfield  {title} {\bibinfo
  {title} {Passive mode locking of thin-disk lasers: effects of spatial hole
  burning},\ }\href {https://doi.org/10.1007/s003400100486} {\bibfield
  {journal} {\bibinfo  {journal} {Appl. Phys. B}\ }\textbf {\bibinfo {volume}
  {72}},\ \bibinfo {pages} {267} (\bibinfo {year} {2001})}\BibitemShut
  {NoStop}%
\bibitem [{\citenamefont {S\"udmeyer}\ \emph {et~al.}(2009)\citenamefont
  {S\"udmeyer}, \citenamefont {Kr\"ankel}, \citenamefont {Baer}, \citenamefont
  {Heckl}, \citenamefont {Saraceno}, \citenamefont {Golling}, \citenamefont
  {Peters}, \citenamefont {Petermann}, \citenamefont {Huber},\ and\
  \citenamefont {Keller}}]{Sudmeyer2009}%
  \BibitemOpen
  \bibfield  {author} {\bibinfo {author} {\bibfnamefont {T.}~\bibnamefont
  {S\"udmeyer}}, \bibinfo {author} {\bibfnamefont {C.}~\bibnamefont
  {Kr\"ankel}}, \bibinfo {author} {\bibfnamefont {C.}~\bibnamefont {Baer}},
  \bibinfo {author} {\bibfnamefont {O.}~\bibnamefont {Heckl}}, \bibinfo
  {author} {\bibfnamefont {C.}~\bibnamefont {Saraceno}}, \bibinfo {author}
  {\bibfnamefont {M.}~\bibnamefont {Golling}}, \bibinfo {author} {\bibfnamefont
  {R.}~\bibnamefont {Peters}}, \bibinfo {author} {\bibfnamefont
  {K.}~\bibnamefont {Petermann}}, \bibinfo {author} {\bibfnamefont
  {G.}~\bibnamefont {Huber}},\ and\ \bibinfo {author} {\bibfnamefont
  {U.}~\bibnamefont {Keller}},\ }\bibfield  {title} {\bibinfo {title}
  {High-power ultrafast thin disk laser oscillators and their potential for
  sub-100-femtosecond pulse generation},\ }\href
  {https://doi.org/10.1007/s00340-009-3700-z} {\bibfield  {journal} {\bibinfo
  {journal} {Appl. Phys. B}\ }\textbf {\bibinfo {volume} {97}},\ \bibinfo
  {pages} {281} (\bibinfo {year} {2009})}\BibitemShut {NoStop}%
\bibitem [{\citenamefont {Baer}\ \emph {et~al.}(2010)\citenamefont {Baer},
  \citenamefont {Kr\"{a}nkel}, \citenamefont {Saraceno}, \citenamefont {Heckl},
  \citenamefont {Golling}, \citenamefont {Peters}, \citenamefont {Petermann},
  \citenamefont {S\"{u}dmeyer}, \citenamefont {Huber},\ and\ \citenamefont
  {Keller}}]{Baer:10}%
  \BibitemOpen
  \bibfield  {author} {\bibinfo {author} {\bibfnamefont {C.~R.~E.}\
  \bibnamefont {Baer}}, \bibinfo {author} {\bibfnamefont {C.}~\bibnamefont
  {Kr\"{a}nkel}}, \bibinfo {author} {\bibfnamefont {C.~J.}\ \bibnamefont
  {Saraceno}}, \bibinfo {author} {\bibfnamefont {O.~H.}\ \bibnamefont {Heckl}},
  \bibinfo {author} {\bibfnamefont {M.}~\bibnamefont {Golling}}, \bibinfo
  {author} {\bibfnamefont {R.}~\bibnamefont {Peters}}, \bibinfo {author}
  {\bibfnamefont {K.}~\bibnamefont {Petermann}}, \bibinfo {author}
  {\bibfnamefont {T.}~\bibnamefont {S\"{u}dmeyer}}, \bibinfo {author}
  {\bibfnamefont {G.}~\bibnamefont {Huber}},\ and\ \bibinfo {author}
  {\bibfnamefont {U.}~\bibnamefont {Keller}},\ }\bibfield  {title} {\bibinfo
  {title} {Femtosecond thin-disk laser with 141 {W} of average power},\ }\href
  {https://doi.org/10.1364/OL.35.002302} {\bibfield  {journal} {\bibinfo
  {journal} {Opt. Lett.}\ }\textbf {\bibinfo {volume} {35}},\ \bibinfo {pages}
  {2302} (\bibinfo {year} {2010})}\BibitemShut {NoStop}%
\bibitem [{\citenamefont {Kaumanns}\ \emph {et~al.}(2021)\citenamefont
  {Kaumanns}, \citenamefont {Kormin}, \citenamefont {Nubbemeyer}, \citenamefont
  {Pervak},\ and\ \citenamefont {Karsch}}]{Kaumanns:21}%
  \BibitemOpen
  \bibfield  {author} {\bibinfo {author} {\bibfnamefont {M.}~\bibnamefont
  {Kaumanns}}, \bibinfo {author} {\bibfnamefont {D.}~\bibnamefont {Kormin}},
  \bibinfo {author} {\bibfnamefont {T.}~\bibnamefont {Nubbemeyer}}, \bibinfo
  {author} {\bibfnamefont {V.}~\bibnamefont {Pervak}},\ and\ \bibinfo {author}
  {\bibfnamefont {S.}~\bibnamefont {Karsch}},\ }\bibfield  {title} {\bibinfo
  {title} {Spectral broadening of 112 {mJ}, 1.3 {ps} pulses at 5 {kHz} in a
  {LG\textsubscript{10}} multipass cell with compressibility to 37{ fs}},\
  }\href {https://doi.org/10.1364/OL.416734} {\bibfield  {journal} {\bibinfo
  {journal} {Opt. Lett.}\ }\textbf {\bibinfo {volume} {46}},\ \bibinfo {pages}
  {929} (\bibinfo {year} {2021})}\BibitemShut {NoStop}%
\bibitem [{\citenamefont {Jakobsson}\ \emph
  {et~al.}(2021{\natexlab{a}})\citenamefont {Jakobsson}, \citenamefont
  {Hooker},\ and\ \citenamefont {Walczak}}]{PhysRevLett.127.184801}%
  \BibitemOpen
  \bibfield  {author} {\bibinfo {author} {\bibfnamefont {O.}~\bibnamefont
  {Jakobsson}}, \bibinfo {author} {\bibfnamefont {S.~M.}\ \bibnamefont
  {Hooker}},\ and\ \bibinfo {author} {\bibfnamefont {R.}~\bibnamefont
  {Walczak}},\ }\bibfield  {title} {\bibinfo {title} {Gev-scale accelerators
  driven by plasma-modulated pulses from kilohertz lasers},\ }\href
  {https://doi.org/10.1103/PhysRevLett.127.184801} {\bibfield  {journal}
  {\bibinfo  {journal} {Phys. Rev. Lett.}\ }\textbf {\bibinfo {volume} {127}},\
  \bibinfo {pages} {184801} (\bibinfo {year} {2021}{\natexlab{a}})}\BibitemShut
  {NoStop}%
\bibitem [{\citenamefont {Eidam}\ \emph {et~al.}(2010)\citenamefont {Eidam},
  \citenamefont {Hanf}, \citenamefont {Seise}, \citenamefont {Andersen},
  \citenamefont {Gabler}, \citenamefont {Wirth}, \citenamefont {Schreiber},
  \citenamefont {Limpert},\ and\ \citenamefont {T\"{u}nnermann}}]{Eidam:10}%
  \BibitemOpen
  \bibfield  {author} {\bibinfo {author} {\bibfnamefont {T.}~\bibnamefont
  {Eidam}}, \bibinfo {author} {\bibfnamefont {S.}~\bibnamefont {Hanf}},
  \bibinfo {author} {\bibfnamefont {E.}~\bibnamefont {Seise}}, \bibinfo
  {author} {\bibfnamefont {T.~V.}\ \bibnamefont {Andersen}}, \bibinfo {author}
  {\bibfnamefont {T.}~\bibnamefont {Gabler}}, \bibinfo {author} {\bibfnamefont
  {C.}~\bibnamefont {Wirth}}, \bibinfo {author} {\bibfnamefont
  {T.}~\bibnamefont {Schreiber}}, \bibinfo {author} {\bibfnamefont
  {J.}~\bibnamefont {Limpert}},\ and\ \bibinfo {author} {\bibfnamefont
  {A.}~\bibnamefont {T\"{u}nnermann}},\ }\bibfield  {title} {\bibinfo {title}
  {Femtosecond fiber {CPA} system emitting 830 {W} average output power},\
  }\href {https://doi.org/10.1364/OL.35.000094} {\bibfield  {journal} {\bibinfo
   {journal} {Opt. Lett.}\ }\textbf {\bibinfo {volume} {35}},\ \bibinfo {pages}
  {94} (\bibinfo {year} {2010})}\BibitemShut {NoStop}%
\bibitem [{\citenamefont {Eidam}\ \emph {et~al.}(2011)\citenamefont {Eidam},
  \citenamefont {Wirth}, \citenamefont {Jauregui}, \citenamefont {Stutzki},
  \citenamefont {Jansen}, \citenamefont {Otto}, \citenamefont {Schmidt},
  \citenamefont {Schreiber}, \citenamefont {Limpert},\ and\ \citenamefont
  {T\"{u}nnermann}}]{Eidam:11}%
  \BibitemOpen
  \bibfield  {author} {\bibinfo {author} {\bibfnamefont {T.}~\bibnamefont
  {Eidam}}, \bibinfo {author} {\bibfnamefont {C.}~\bibnamefont {Wirth}},
  \bibinfo {author} {\bibfnamefont {C.}~\bibnamefont {Jauregui}}, \bibinfo
  {author} {\bibfnamefont {F.}~\bibnamefont {Stutzki}}, \bibinfo {author}
  {\bibfnamefont {F.}~\bibnamefont {Jansen}}, \bibinfo {author} {\bibfnamefont
  {H.-J.}\ \bibnamefont {Otto}}, \bibinfo {author} {\bibfnamefont
  {O.}~\bibnamefont {Schmidt}}, \bibinfo {author} {\bibfnamefont
  {T.}~\bibnamefont {Schreiber}}, \bibinfo {author} {\bibfnamefont
  {J.}~\bibnamefont {Limpert}},\ and\ \bibinfo {author} {\bibfnamefont
  {A.}~\bibnamefont {T\"{u}nnermann}},\ }\bibfield  {title} {\bibinfo {title}
  {Experimental observations of the threshold-like onset of mode instabilities
  in high power fiber amplifiers},\ }\href
  {https://doi.org/10.1364/OE.19.013218} {\bibfield  {journal} {\bibinfo
  {journal} {Opt. Express}\ }\textbf {\bibinfo {volume} {19}},\ \bibinfo
  {pages} {13218} (\bibinfo {year} {2011})}\BibitemShut {NoStop}%
\bibitem [{\citenamefont {Jauregui}\ \emph {et~al.}(2011)\citenamefont
  {Jauregui}, \citenamefont {Eidam}, \citenamefont {Limpert},\ and\
  \citenamefont {T\"{u}nnermann}}]{Jauregui:11}%
  \BibitemOpen
  \bibfield  {author} {\bibinfo {author} {\bibfnamefont {C.}~\bibnamefont
  {Jauregui}}, \bibinfo {author} {\bibfnamefont {T.}~\bibnamefont {Eidam}},
  \bibinfo {author} {\bibfnamefont {J.}~\bibnamefont {Limpert}},\ and\ \bibinfo
  {author} {\bibfnamefont {A.}~\bibnamefont {T\"{u}nnermann}},\ }\bibfield
  {title} {\bibinfo {title} {Impact of modal interference on the beam quality
  of high-power fiber amplifiers},\ }\href
  {https://doi.org/10.1364/OE.19.003258} {\bibfield  {journal} {\bibinfo
  {journal} {Opt. Express}\ }\textbf {\bibinfo {volume} {19}},\ \bibinfo
  {pages} {3258} (\bibinfo {year} {2011})}\BibitemShut {NoStop}%
\bibitem [{\citenamefont {Smith}\ and\ \citenamefont {Smith}(2011)}]{Smith:11}%
  \BibitemOpen
  \bibfield  {author} {\bibinfo {author} {\bibfnamefont {A.~V.}\ \bibnamefont
  {Smith}}\ and\ \bibinfo {author} {\bibfnamefont {J.~J.}\ \bibnamefont
  {Smith}},\ }\bibfield  {title} {\bibinfo {title} {Mode instability in high
  power fiber amplifiers},\ }\href {https://doi.org/10.1364/OE.19.010180}
  {\bibfield  {journal} {\bibinfo  {journal} {Opt. Express}\ }\textbf {\bibinfo
  {volume} {19}},\ \bibinfo {pages} {10180} (\bibinfo {year}
  {2011})}\BibitemShut {NoStop}%
\bibitem [{\citenamefont {Jauregui}\ \emph {et~al.}(2020)\citenamefont
  {Jauregui}, \citenamefont {Stihler},\ and\ \citenamefont
  {Limpert}}]{Jauregui:20}%
  \BibitemOpen
  \bibfield  {author} {\bibinfo {author} {\bibfnamefont {C.}~\bibnamefont
  {Jauregui}}, \bibinfo {author} {\bibfnamefont {C.}~\bibnamefont {Stihler}},\
  and\ \bibinfo {author} {\bibfnamefont {J.}~\bibnamefont {Limpert}},\
  }\bibfield  {title} {\bibinfo {title} {Transverse mode instability},\ }\href
  {https://doi.org/10.1364/AOP.385184} {\bibfield  {journal} {\bibinfo
  {journal} {Adv. Opt. Photon.}\ }\textbf {\bibinfo {volume} {12}},\ \bibinfo
  {pages} {429} (\bibinfo {year} {2020})}\BibitemShut {NoStop}%
\bibitem [{\citenamefont {Esarey}\ \emph {et~al.}(1993)\citenamefont {Esarey},
  \citenamefont {Sprangle}, \citenamefont {Krall}, \citenamefont {Ting},\ and\
  \citenamefont {Joyce}}]{doi:10.1063/1.860707}%
  \BibitemOpen
  \bibfield  {author} {\bibinfo {author} {\bibfnamefont {E.}~\bibnamefont
  {Esarey}}, \bibinfo {author} {\bibfnamefont {P.}~\bibnamefont {Sprangle}},
  \bibinfo {author} {\bibfnamefont {J.}~\bibnamefont {Krall}}, \bibinfo
  {author} {\bibfnamefont {A.}~\bibnamefont {Ting}},\ and\ \bibinfo {author}
  {\bibfnamefont {G.}~\bibnamefont {Joyce}},\ }\bibfield  {title} {\bibinfo
  {title} {Optically guided laser wake‐field acceleration*},\ }\href
  {https://doi.org/10.1063/1.860707} {\bibfield  {journal} {\bibinfo  {journal}
  {Physics of Fluids B: Plasma Physics}\ }\textbf {\bibinfo {volume} {5}},\
  \bibinfo {pages} {2690} (\bibinfo {year} {1993})},\ \Eprint
  {https://arxiv.org/abs/https://doi.org/10.1063/1.860707}
  {https://doi.org/10.1063/1.860707} \BibitemShut {NoStop}%
\bibitem [{\citenamefont {Andreev}\ \emph {et~al.}(1994)\citenamefont
  {Andreev}, \citenamefont {Gorbunov}, \citenamefont {Kirsanov}, \citenamefont
  {Pogosova},\ and\ \citenamefont {Ramazashvili}}]{NEAndreev_1994}%
  \BibitemOpen
  \bibfield  {author} {\bibinfo {author} {\bibfnamefont {N.~E.}\ \bibnamefont
  {Andreev}}, \bibinfo {author} {\bibfnamefont {L.~M.}\ \bibnamefont
  {Gorbunov}}, \bibinfo {author} {\bibfnamefont {V.~I.}\ \bibnamefont
  {Kirsanov}}, \bibinfo {author} {\bibfnamefont {A.~A.}\ \bibnamefont
  {Pogosova}},\ and\ \bibinfo {author} {\bibfnamefont {R.~R.}\ \bibnamefont
  {Ramazashvili}},\ }\bibfield  {title} {\bibinfo {title} {The theory of laser
  self-resonant wake field excitation},\ }\href
  {https://doi.org/10.1088/0031-8949/49/1/014} {\bibfield  {journal} {\bibinfo
  {journal} {Physica Scripta}\ }\textbf {\bibinfo {volume} {49}},\ \bibinfo
  {pages} {101} (\bibinfo {year} {1994})}\BibitemShut {NoStop}%
\bibitem [{\citenamefont {Mora}\ and\ \citenamefont
  {Antonsen}(1997)}]{doi:10.1063/1.872134}%
  \BibitemOpen
  \bibfield  {author} {\bibinfo {author} {\bibfnamefont {P.}~\bibnamefont
  {Mora}}\ and\ \bibinfo {author} {\bibfnamefont {T.~M.}\ \bibnamefont
  {Antonsen}, \bibfnamefont {Jr.}},\ }\bibfield  {title} {\bibinfo {title}
  {Kinetic modeling of intense, short laser pulses propagating in tenuous
  plasmas},\ }\href {https://doi.org/10.1063/1.872134} {\bibfield  {journal}
  {\bibinfo  {journal} {Physics of Plasmas}\ }\textbf {\bibinfo {volume} {4}},\
  \bibinfo {pages} {217} (\bibinfo {year} {1997})},\ \Eprint
  {https://arxiv.org/abs/https://doi.org/10.1063/1.872134}
  {https://doi.org/10.1063/1.872134} \BibitemShut {NoStop}%
\bibitem [{\citenamefont {Zhu}\ \emph {et~al.}(2012)\citenamefont {Zhu},
  \citenamefont {Palastro},\ and\ \citenamefont
  {Antonsen}}]{doi:10.1063/1.3691837}%
  \BibitemOpen
  \bibfield  {author} {\bibinfo {author} {\bibfnamefont {W.}~\bibnamefont
  {Zhu}}, \bibinfo {author} {\bibfnamefont {J.~P.}\ \bibnamefont {Palastro}},\
  and\ \bibinfo {author} {\bibfnamefont {T.~M.}\ \bibnamefont {Antonsen}},\
  }\bibfield  {title} {\bibinfo {title} {Studies of spectral modification and
  limitations of the modified paraxial equation in laser wakefield
  simulations},\ }\href {https://doi.org/10.1063/1.3691837} {\bibfield
  {journal} {\bibinfo  {journal} {Physics of Plasmas}\ }\textbf {\bibinfo
  {volume} {19}},\ \bibinfo {pages} {033105} (\bibinfo {year} {2012})},\
  \Eprint {https://arxiv.org/abs/https://doi.org/10.1063/1.3691837}
  {https://doi.org/10.1063/1.3691837} \BibitemShut {NoStop}%
\bibitem [{sup()}]{supp}%
  \BibitemOpen
  \href@noop {} {}\bibinfo {howpublished}
  {\url{URL_will_be_inserted_by_publisher}}\BibitemShut {NoStop}%
\bibitem [{\citenamefont {Esarey}\ \emph {et~al.}(1995)\citenamefont {Esarey},
  \citenamefont {Sprangle}, \citenamefont {Pilloff},\ and\ \citenamefont
  {Krall}}]{Esarey:95}%
  \BibitemOpen
  \bibfield  {author} {\bibinfo {author} {\bibfnamefont {E.}~\bibnamefont
  {Esarey}}, \bibinfo {author} {\bibfnamefont {P.}~\bibnamefont {Sprangle}},
  \bibinfo {author} {\bibfnamefont {M.}~\bibnamefont {Pilloff}},\ and\ \bibinfo
  {author} {\bibfnamefont {J.}~\bibnamefont {Krall}},\ }\bibfield  {title}
  {\bibinfo {title} {Theory and group velocity of ultrashort, tightly focused
  laser pulses},\ }\href {https://doi.org/10.1364/JOSAB.12.001695} {\bibfield
  {journal} {\bibinfo  {journal} {J. Opt. Soc. Am. B}\ }\textbf {\bibinfo
  {volume} {12}},\ \bibinfo {pages} {1695} (\bibinfo {year}
  {1995})}\BibitemShut {NoStop}%
\bibitem [{\citenamefont {Esarey}\ and\ \citenamefont
  {Leemans}(1999)}]{PhysRevE.59.1082}%
  \BibitemOpen
  \bibfield  {author} {\bibinfo {author} {\bibfnamefont {E.}~\bibnamefont
  {Esarey}}\ and\ \bibinfo {author} {\bibfnamefont {W.~P.}\ \bibnamefont
  {Leemans}},\ }\bibfield  {title} {\bibinfo {title} {Nonparaxial propagation
  of ultrashort laser pulses in plasma channels},\ }\href
  {https://doi.org/10.1103/PhysRevE.59.1082} {\bibfield  {journal} {\bibinfo
  {journal} {Phys. Rev. E}\ }\textbf {\bibinfo {volume} {59}},\ \bibinfo
  {pages} {1082} (\bibinfo {year} {1999})}\BibitemShut {NoStop}%
\bibitem [{\citenamefont {Kurtz}\ and\ \citenamefont
  {Streifer}(1969)}]{1126872}%
  \BibitemOpen
  \bibfield  {author} {\bibinfo {author} {\bibfnamefont {C.}~\bibnamefont
  {Kurtz}}\ and\ \bibinfo {author} {\bibfnamefont {W.}~\bibnamefont
  {Streifer}},\ }\bibfield  {title} {\bibinfo {title} {Guided waves in
  inhomogeneous focusing media part i: Formulation, solution for quadratic
  inhomogeneity},\ }\href {https://doi.org/10.1109/TMTT.1969.1126872}
  {\bibfield  {journal} {\bibinfo  {journal} {IEEE Transactions on Microwave
  Theory and Techniques}\ }\textbf {\bibinfo {volume} {17}},\ \bibinfo {pages}
  {11} (\bibinfo {year} {1969})}\BibitemShut {NoStop}%
\bibitem [{\citenamefont {Firth}(1977)}]{FIRTH1977226}%
  \BibitemOpen
  \bibfield  {author} {\bibinfo {author} {\bibfnamefont {W.}~\bibnamefont
  {Firth}},\ }\bibfield  {title} {\bibinfo {title} {Propagation of laser beams
  through inhomogeneous media},\ }\href
  {https://doi.org/https://doi.org/10.1016/0030-4018(77)90024-4} {\bibfield
  {journal} {\bibinfo  {journal} {Optics Communications}\ }\textbf {\bibinfo
  {volume} {22}},\ \bibinfo {pages} {226} (\bibinfo {year} {1977})}\BibitemShut
  {NoStop}%
\bibitem [{\citenamefont {Jakobsson}\ \emph
  {et~al.}(2021{\natexlab{b}})\citenamefont {Jakobsson}, \citenamefont
  {Hooker},\ and\ \citenamefont {Walczak}}]{PhysRevLett.127.184801_supp}%
  \BibitemOpen
  \bibfield  {author} {\bibinfo {author} {\bibfnamefont {O.}~\bibnamefont
  {Jakobsson}}, \bibinfo {author} {\bibfnamefont {S.~M.}\ \bibnamefont
  {Hooker}},\ and\ \bibinfo {author} {\bibfnamefont {R.}~\bibnamefont
  {Walczak}},\ }\href@noop {} {\bibinfo {title} {Supplemental material for
  gev-scale accelerators driven by plasma-modulated pulses from kilohertz
  lasers}},\ \bibinfo {howpublished}
  {\url{http://link.aps.org/supplemental/10.1103/PhysRevLett.127.184801}}
  (\bibinfo {year} {2021}{\natexlab{b}})\BibitemShut {NoStop}%
\bibitem [{\citenamefont {Gibbon}(1990)}]{doi:10.1063/1.859401}%
  \BibitemOpen
  \bibfield  {author} {\bibinfo {author} {\bibfnamefont {P.}~\bibnamefont
  {Gibbon}},\ }\bibfield  {title} {\bibinfo {title} {The self‐trapping of
  light waves by beat‐wave excitation.},\ }\href
  {https://doi.org/10.1063/1.859401} {\bibfield  {journal} {\bibinfo  {journal}
  {Physics of Fluids B: Plasma Physics}\ }\textbf {\bibinfo {volume} {2}},\
  \bibinfo {pages} {2196} (\bibinfo {year} {1990})},\ \Eprint
  {https://arxiv.org/abs/https://doi.org/10.1063/1.859401}
  {https://doi.org/10.1063/1.859401} \BibitemShut {NoStop}%
\bibitem [{\citenamefont {Mori}(1997)}]{641309}%
  \BibitemOpen
  \bibfield  {author} {\bibinfo {author} {\bibfnamefont {W.}~\bibnamefont
  {Mori}},\ }\bibfield  {title} {\bibinfo {title} {The physics of the nonlinear
  optics of plasmas at relativistic intensities for short-pulse lasers},\
  }\href {https://doi.org/10.1109/3.641309} {\bibfield  {journal} {\bibinfo
  {journal} {IEEE Journal of Quantum Electronics}\ }\textbf {\bibinfo {volume}
  {33}},\ \bibinfo {pages} {1942} (\bibinfo {year} {1997})}\BibitemShut
  {NoStop}%
\bibitem [{\citenamefont {Max}\ \emph {et~al.}(1974)\citenamefont {Max},
  \citenamefont {Arons},\ and\ \citenamefont {Langdon}}]{PhysRevLett.33.209}%
  \BibitemOpen
  \bibfield  {author} {\bibinfo {author} {\bibfnamefont {C.~E.}\ \bibnamefont
  {Max}}, \bibinfo {author} {\bibfnamefont {J.}~\bibnamefont {Arons}},\ and\
  \bibinfo {author} {\bibfnamefont {A.~B.}\ \bibnamefont {Langdon}},\
  }\bibfield  {title} {\bibinfo {title} {Self-modulation and self-focusing of
  electromagnetic waves in plasmas},\ }\href
  {https://doi.org/10.1103/PhysRevLett.33.209} {\bibfield  {journal} {\bibinfo
  {journal} {Phys. Rev. Lett.}\ }\textbf {\bibinfo {volume} {33}},\ \bibinfo
  {pages} {209} (\bibinfo {year} {1974})}\BibitemShut {NoStop}%
\bibitem [{\citenamefont {Esarey}\ \emph {et~al.}(1997)\citenamefont {Esarey},
  \citenamefont {Sprangle}, \citenamefont {Krall},\ and\ \citenamefont
  {Ting}}]{641305}%
  \BibitemOpen
  \bibfield  {author} {\bibinfo {author} {\bibfnamefont {E.}~\bibnamefont
  {Esarey}}, \bibinfo {author} {\bibfnamefont {P.}~\bibnamefont {Sprangle}},
  \bibinfo {author} {\bibfnamefont {J.}~\bibnamefont {Krall}},\ and\ \bibinfo
  {author} {\bibfnamefont {A.}~\bibnamefont {Ting}},\ }\bibfield  {title}
  {\bibinfo {title} {Self-focusing and guiding of short laser pulses in
  ionizing gases and plasmas},\ }\href {https://doi.org/10.1109/3.641305}
  {\bibfield  {journal} {\bibinfo  {journal} {IEEE Journal of Quantum
  Electronics}\ }\textbf {\bibinfo {volume} {33}},\ \bibinfo {pages} {1879}
  (\bibinfo {year} {1997})}\BibitemShut {NoStop}%
\bibitem [{\citenamefont {Gonsalves}\ \emph {et~al.}(2010)\citenamefont
  {Gonsalves}, \citenamefont {Nakamura}, \citenamefont {Lin}, \citenamefont
  {Osterhoff}, \citenamefont {Shiraishi}, \citenamefont {Schroeder},
  \citenamefont {Geddes}, \citenamefont {Tóth}, \citenamefont {Esarey},\ and\
  \citenamefont {Leemans}}]{doi:10.1063/1.3357175}%
  \BibitemOpen
  \bibfield  {author} {\bibinfo {author} {\bibfnamefont {A.~J.}\ \bibnamefont
  {Gonsalves}}, \bibinfo {author} {\bibfnamefont {K.}~\bibnamefont {Nakamura}},
  \bibinfo {author} {\bibfnamefont {C.}~\bibnamefont {Lin}}, \bibinfo {author}
  {\bibfnamefont {J.}~\bibnamefont {Osterhoff}}, \bibinfo {author}
  {\bibfnamefont {S.}~\bibnamefont {Shiraishi}}, \bibinfo {author}
  {\bibfnamefont {C.~B.}\ \bibnamefont {Schroeder}}, \bibinfo {author}
  {\bibfnamefont {C.~G.~R.}\ \bibnamefont {Geddes}}, \bibinfo {author}
  {\bibfnamefont {C.}~\bibnamefont {Tóth}}, \bibinfo {author} {\bibfnamefont
  {E.}~\bibnamefont {Esarey}},\ and\ \bibinfo {author} {\bibfnamefont {W.~P.}\
  \bibnamefont {Leemans}},\ }\bibfield  {title} {\bibinfo {title} {Plasma
  channel diagnostic based on laser centroid oscillations},\ }\href
  {https://doi.org/10.1063/1.3357175} {\bibfield  {journal} {\bibinfo
  {journal} {Physics of Plasmas}\ }\textbf {\bibinfo {volume} {17}},\ \bibinfo
  {pages} {056706} (\bibinfo {year} {2010})},\ \Eprint
  {https://arxiv.org/abs/https://doi.org/10.1063/1.3357175}
  {https://doi.org/10.1063/1.3357175} \BibitemShut {NoStop}%
\bibitem [{\citenamefont {Andreev}\ \emph {et~al.}(1997)\citenamefont
  {Andreev}, \citenamefont {Gorbunov}, \citenamefont {Kirsanov}, \citenamefont
  {Nakajima},\ and\ \citenamefont {Ogata}}]{doi:10.1063/1.872186}%
  \BibitemOpen
  \bibfield  {author} {\bibinfo {author} {\bibfnamefont {N.~E.}\ \bibnamefont
  {Andreev}}, \bibinfo {author} {\bibfnamefont {L.~M.}\ \bibnamefont
  {Gorbunov}}, \bibinfo {author} {\bibfnamefont {V.~I.}\ \bibnamefont
  {Kirsanov}}, \bibinfo {author} {\bibfnamefont {K.}~\bibnamefont {Nakajima}},\
  and\ \bibinfo {author} {\bibfnamefont {A.}~\bibnamefont {Ogata}},\ }\bibfield
   {title} {\bibinfo {title} {Structure of the wake field in plasma channels},\
  }\href {https://doi.org/10.1063/1.872186} {\bibfield  {journal} {\bibinfo
  {journal} {Physics of Plasmas}\ }\textbf {\bibinfo {volume} {4}},\ \bibinfo
  {pages} {1145} (\bibinfo {year} {1997})},\ \Eprint
  {https://arxiv.org/abs/https://doi.org/10.1063/1.872186}
  {https://doi.org/10.1063/1.872186} \BibitemShut {NoStop}%
\bibitem [{\citenamefont {Rosenbluth}\ and\ \citenamefont
  {Liu}(1972)}]{PhysRevLett.29.701}%
  \BibitemOpen
  \bibfield  {author} {\bibinfo {author} {\bibfnamefont {M.~N.}\ \bibnamefont
  {Rosenbluth}}\ and\ \bibinfo {author} {\bibfnamefont {C.~S.}\ \bibnamefont
  {Liu}},\ }\bibfield  {title} {\bibinfo {title} {Excitation of plasma waves by
  two laser beams},\ }\href {https://doi.org/10.1103/PhysRevLett.29.701}
  {\bibfield  {journal} {\bibinfo  {journal} {Phys. Rev. Lett.}\ }\textbf
  {\bibinfo {volume} {29}},\ \bibinfo {pages} {701} (\bibinfo {year}
  {1972})}\BibitemShut {NoStop}%
\bibitem [{\citenamefont {Siegrist}(1976)}]{SIEGRIST1976402}%
  \BibitemOpen
  \bibfield  {author} {\bibinfo {author} {\bibfnamefont {M.}~\bibnamefont
  {Siegrist}},\ }\bibfield  {title} {\bibinfo {title} {Self-focusing in a
  plasma due to ponderomotive forces and relativistic effects},\ }\href
  {https://doi.org/https://doi.org/10.1016/0030-4018(76)90028-6} {\bibfield
  {journal} {\bibinfo  {journal} {Optics Communications}\ }\textbf {\bibinfo
  {volume} {16}},\ \bibinfo {pages} {402} (\bibinfo {year} {1976})}\BibitemShut
  {NoStop}%
\bibitem [{\citenamefont {Brandi}\ \emph {et~al.}(1993)\citenamefont {Brandi},
  \citenamefont {Manus}, \citenamefont {Mainfray}, \citenamefont {Lehner},\
  and\ \citenamefont {Bonnaud}}]{doi:10.1063/1.860828}%
  \BibitemOpen
  \bibfield  {author} {\bibinfo {author} {\bibfnamefont {H.~S.}\ \bibnamefont
  {Brandi}}, \bibinfo {author} {\bibfnamefont {C.}~\bibnamefont {Manus}},
  \bibinfo {author} {\bibfnamefont {G.}~\bibnamefont {Mainfray}}, \bibinfo
  {author} {\bibfnamefont {T.}~\bibnamefont {Lehner}},\ and\ \bibinfo {author}
  {\bibfnamefont {G.}~\bibnamefont {Bonnaud}},\ }\bibfield  {title} {\bibinfo
  {title} {Relativistic and ponderomotive self‐focusing of a laser beam in a
  radially inhomogeneous plasma. i. paraxial approximation},\ }\href
  {https://doi.org/10.1063/1.860828} {\bibfield  {journal} {\bibinfo  {journal}
  {Physics of Fluids B: Plasma Physics}\ }\textbf {\bibinfo {volume} {5}},\
  \bibinfo {pages} {3539} (\bibinfo {year} {1993})},\ \Eprint
  {https://arxiv.org/abs/https://doi.org/10.1063/1.860828}
  {https://doi.org/10.1063/1.860828} \BibitemShut {NoStop}%
\bibitem [{\citenamefont {Djordjević}\ \emph {et~al.}(2017)\citenamefont
  {Djordjević}, \citenamefont {Benedetti}, \citenamefont {Schroeder},
  \citenamefont {Esarey},\ and\ \citenamefont
  {Leemans}}]{doi:10.1063/1.4975860}%
  \BibitemOpen
  \bibfield  {author} {\bibinfo {author} {\bibfnamefont {B.~Z.}\ \bibnamefont
  {Djordjević}}, \bibinfo {author} {\bibfnamefont {C.}~\bibnamefont
  {Benedetti}}, \bibinfo {author} {\bibfnamefont {C.~B.}\ \bibnamefont
  {Schroeder}}, \bibinfo {author} {\bibfnamefont {E.}~\bibnamefont {Esarey}},\
  and\ \bibinfo {author} {\bibfnamefont {W.~P.}\ \bibnamefont {Leemans}},\
  }\bibfield  {title} {\bibinfo {title} {Laser mode control using leaky plasma
  channels},\ }\href {https://doi.org/10.1063/1.4975860} {\bibfield  {journal}
  {\bibinfo  {journal} {AIP Conference Proceedings}\ }\textbf {\bibinfo
  {volume} {1812}},\ \bibinfo {pages} {040013} (\bibinfo {year} {2017})},\
  \Eprint
  {https://arxiv.org/abs/https://aip.scitation.org/doi/pdf/10.1063/1.4975860}
  {https://aip.scitation.org/doi/pdf/10.1063/1.4975860} \BibitemShut {NoStop}%
\bibitem [{\citenamefont {Djordjević}\ \emph {et~al.}(2018)\citenamefont
  {Djordjević}, \citenamefont {Benedetti}, \citenamefont {Schroeder},
  \citenamefont {Esarey},\ and\ \citenamefont
  {Leemans}}]{doi:10.1063/1.5006198}%
  \BibitemOpen
  \bibfield  {author} {\bibinfo {author} {\bibfnamefont {B.~Z.}\ \bibnamefont
  {Djordjević}}, \bibinfo {author} {\bibfnamefont {C.}~\bibnamefont
  {Benedetti}}, \bibinfo {author} {\bibfnamefont {C.~B.}\ \bibnamefont
  {Schroeder}}, \bibinfo {author} {\bibfnamefont {E.}~\bibnamefont {Esarey}},\
  and\ \bibinfo {author} {\bibfnamefont {W.~P.}\ \bibnamefont {Leemans}},\
  }\bibfield  {title} {\bibinfo {title} {Filtering higher-order laser modes
  using leaky plasma channels},\ }\href {https://doi.org/10.1063/1.5006198}
  {\bibfield  {journal} {\bibinfo  {journal} {Physics of Plasmas}\ }\textbf
  {\bibinfo {volume} {25}},\ \bibinfo {pages} {013103} (\bibinfo {year}
  {2018})},\ \Eprint {https://arxiv.org/abs/https://doi.org/10.1063/1.5006198}
  {https://doi.org/10.1063/1.5006198} \BibitemShut {NoStop}%
\bibitem [{\citenamefont {Fedeli}\ \emph {et~al.}(2022)\citenamefont {Fedeli},
  \citenamefont {Huebl}, \citenamefont {Boillod-Cerneux}, \citenamefont
  {Clark}, \citenamefont {Gott}, \citenamefont {Hillairet}, \citenamefont
  {Jaure}, \citenamefont {Leblanc}, \citenamefont {Lehe}, \citenamefont
  {Myers}, \citenamefont {Piechurski}, \citenamefont {Sato}, \citenamefont
  {Zaim}, \citenamefont {Zhang}, \citenamefont {Vay},\ and\ \citenamefont
  {Vincenti}}]{WarpX}%
  \BibitemOpen
  \bibfield  {author} {\bibinfo {author} {\bibfnamefont {L.}~\bibnamefont
  {Fedeli}}, \bibinfo {author} {\bibfnamefont {A.}~\bibnamefont {Huebl}},
  \bibinfo {author} {\bibfnamefont {F.}~\bibnamefont {Boillod-Cerneux}},
  \bibinfo {author} {\bibfnamefont {T.}~\bibnamefont {Clark}}, \bibinfo
  {author} {\bibfnamefont {K.}~\bibnamefont {Gott}}, \bibinfo {author}
  {\bibfnamefont {C.}~\bibnamefont {Hillairet}}, \bibinfo {author}
  {\bibfnamefont {S.}~\bibnamefont {Jaure}}, \bibinfo {author} {\bibfnamefont
  {A.}~\bibnamefont {Leblanc}}, \bibinfo {author} {\bibfnamefont
  {R.}~\bibnamefont {Lehe}}, \bibinfo {author} {\bibfnamefont {A.}~\bibnamefont
  {Myers}}, \bibinfo {author} {\bibfnamefont {C.}~\bibnamefont {Piechurski}},
  \bibinfo {author} {\bibfnamefont {M.}~\bibnamefont {Sato}}, \bibinfo {author}
  {\bibfnamefont {N.}~\bibnamefont {Zaim}}, \bibinfo {author} {\bibfnamefont
  {W.}~\bibnamefont {Zhang}}, \bibinfo {author} {\bibfnamefont
  {J.}~\bibnamefont {Vay}},\ and\ \bibinfo {author} {\bibfnamefont
  {H.}~\bibnamefont {Vincenti}},\ }\bibfield  {title} {\bibinfo {title}
  {Pushing the frontier in the design of laser-based electron accelerators with
  groundbreaking mesh-refined particle-in-cell simulations on exascale-class
  supercomputers},\ }in\ \href {https://doi.ieeecomputersociety.org/} {\emph
  {\bibinfo {booktitle} {2022 SC22: International Conference for High
  Performance Computing, Networking, Storage and Analysis (SC) (SC)}}}\
  (\bibinfo  {publisher} {IEEE Computer Society},\ \bibinfo {address} {Los
  Alamitos, CA, USA},\ \bibinfo {year} {2022})\ pp.\ \bibinfo {pages}
  {25--36}\BibitemShut {NoStop}%
\end{thebibliography}%


%apsrev4-2.bst 2019-01-14 (MD) hand-edited version of apsrev4-1.bst
%Control: key (0)
%Control: author (8) initials jnrlst
%Control: editor formatted (1) identically to author
%Control: production of article title (0) allowed
%Control: page (0) single
%Control: year (1) truncated
%Control: production of eprint (0) enabled
\providecommand{\noopsort}[1]{}\providecommand{\singleletter}[1]{#1}%
\begin{thebibliography}{5}%
\makeatletter
\providecommand \@ifxundefined [1]{%
 \@ifx{#1\undefined}
}%
\providecommand \@ifnum [1]{%
 \ifnum #1\expandafter \@firstoftwo
 \else \expandafter \@secondoftwo
 \fi
}%
\providecommand \@ifx [1]{%
 \ifx #1\expandafter \@firstoftwo
 \else \expandafter \@secondoftwo
 \fi
}%
\providecommand \natexlab [1]{#1}%
\providecommand \enquote  [1]{``#1''}%
\providecommand \bibnamefont  [1]{#1}%
\providecommand \bibfnamefont [1]{#1}%
\providecommand \citenamefont [1]{#1}%
\providecommand \href@noop [0]{\@secondoftwo}%
\providecommand \href [0]{\begingroup \@sanitize@url \@href}%
\providecommand \@href[1]{\@@startlink{#1}\@@href}%
\providecommand \@@href[1]{\endgroup#1\@@endlink}%
\providecommand \@sanitize@url [0]{\catcode `\\12\catcode `\$12\catcode
  `\&12\catcode `\#12\catcode `\^12\catcode `\_12\catcode `\%12\relax}%
\providecommand \@@startlink[1]{}%
\providecommand \@@endlink[0]{}%
\providecommand \url  [0]{\begingroup\@sanitize@url \@url }%
\providecommand \@url [1]{\endgroup\@href {#1}{\urlprefix }}%
\providecommand \urlprefix  [0]{URL }%
\providecommand \Eprint [0]{\href }%
\providecommand \doibase [0]{https://doi.org/}%
\providecommand \selectlanguage [0]{\@gobble}%
\providecommand \bibinfo  [0]{\@secondoftwo}%
\providecommand \bibfield  [0]{\@secondoftwo}%
\providecommand \translation [1]{[#1]}%
\providecommand \BibitemOpen [0]{}%
\providecommand \bibitemStop [0]{}%
\providecommand \bibitemNoStop [0]{.\EOS\space}%
\providecommand \EOS [0]{\spacefactor3000\relax}%
\providecommand \BibitemShut  [1]{\csname bibitem#1\endcsname}%
\let\auto@bib@innerbib\@empty
%</preamble>
\bibitem [{\citenamefont {Esarey}\ \emph {et~al.}(1995)\citenamefont {Esarey},
  \citenamefont {Sprangle}, \citenamefont {Pilloff},\ and\ \citenamefont
  {Krall}}]{SpotVG}%
  \BibitemOpen
  \bibfield  {author} {\bibinfo {author} {\bibfnamefont {E.}~\bibnamefont
  {Esarey}}, \bibinfo {author} {\bibfnamefont {P.}~\bibnamefont {Sprangle}},
  \bibinfo {author} {\bibfnamefont {M.}~\bibnamefont {Pilloff}},\ and\ \bibinfo
  {author} {\bibfnamefont {J.}~\bibnamefont {Krall}},\ }\bibfield  {title}
  {\bibinfo {title} {Theory and group velocity of ultrashort, tightly focused
  laser pulses},\ }\href {https://doi.org/10.1364/JOSAB.12.001695} {\bibfield
  {journal} {\bibinfo  {journal} {J. Opt. Soc. Am. B}\ }\textbf {\bibinfo
  {volume} {12}},\ \bibinfo {pages} {1695} (\bibinfo {year}
  {1995})}\BibitemShut {NoStop}%
\bibitem [{\citenamefont {Jakobsson}\ \emph {et~al.}(2021)\citenamefont
  {Jakobsson}, \citenamefont {Hooker},\ and\ \citenamefont
  {Walczak}}]{PhysRevLett.127.184801}%
  \BibitemOpen
  \bibfield  {author} {\bibinfo {author} {\bibfnamefont {O.}~\bibnamefont
  {Jakobsson}}, \bibinfo {author} {\bibfnamefont {S.~M.}\ \bibnamefont
  {Hooker}},\ and\ \bibinfo {author} {\bibfnamefont {R.}~\bibnamefont
  {Walczak}},\ }\bibfield  {title} {\bibinfo {title} {Gev-scale accelerators
  driven by plasma-modulated pulses from kilohertz lasers},\ }\href
  {https://doi.org/10.1103/PhysRevLett.127.184801} {\bibfield  {journal}
  {\bibinfo  {journal} {Phys. Rev. Lett.}\ }\textbf {\bibinfo {volume} {127}},\
  \bibinfo {pages} {184801} (\bibinfo {year} {2021})}\BibitemShut {NoStop}%
\bibitem [{\citenamefont {Esarey}\ and\ \citenamefont
  {Leemans}(1999)}]{SpotVGplasma}%
  \BibitemOpen
  \bibfield  {author} {\bibinfo {author} {\bibfnamefont {E.}~\bibnamefont
  {Esarey}}\ and\ \bibinfo {author} {\bibfnamefont {W.~P.}\ \bibnamefont
  {Leemans}},\ }\bibfield  {title} {\bibinfo {title} {Nonparaxial propagation
  of ultrashort laser pulses in plasma channels},\ }\href
  {https://doi.org/10.1103/PhysRevE.59.1082} {\bibfield  {journal} {\bibinfo
  {journal} {Phys. Rev. E}\ }\textbf {\bibinfo {volume} {59}},\ \bibinfo
  {pages} {1082} (\bibinfo {year} {1999})}\BibitemShut {NoStop}%
\bibitem [{\citenamefont {Andreev}\ \emph {et~al.}(1996)\citenamefont {Andreev}
  \emph {et~al.}}]{Andreev96}%
  \BibitemOpen
  \bibfield  {author} {\bibinfo {author} {\bibfnamefont {N.~E.}\ \bibnamefont
  {Andreev}} \emph {et~al.},\ }\bibfield  {title} {\bibinfo {title} {Structure
  of the wake field in plasma channels},\ }\href@noop {} {\bibfield  {journal}
  {\bibinfo  {journal} {Phys. Plasmas}\ }\textbf {\bibinfo {volume} {4}},\
  \bibinfo {pages} {1145} (\bibinfo {year} {1996})}\BibitemShut {NoStop}%
\bibitem [{\citenamefont {Fedeli}\ \emph {et~al.}(2022)\citenamefont {Fedeli},
  \citenamefont {Huebl}, \citenamefont {Boillod-Cerneux}, \citenamefont
  {Clark}, \citenamefont {Gott}, \citenamefont {Hillairet}, \citenamefont
  {Jaure}, \citenamefont {Leblanc}, \citenamefont {Lehe}, \citenamefont
  {Myers}, \citenamefont {Piechurski}, \citenamefont {Sato}, \citenamefont
  {Zaim}, \citenamefont {Zhang}, \citenamefont {Vay},\ and\ \citenamefont
  {Vincenti}}]{WarpX}%
  \BibitemOpen
  \bibfield  {author} {\bibinfo {author} {\bibfnamefont {L.}~\bibnamefont
  {Fedeli}}, \bibinfo {author} {\bibfnamefont {A.}~\bibnamefont {Huebl}},
  \bibinfo {author} {\bibfnamefont {F.}~\bibnamefont {Boillod-Cerneux}},
  \bibinfo {author} {\bibfnamefont {T.}~\bibnamefont {Clark}}, \bibinfo
  {author} {\bibfnamefont {K.}~\bibnamefont {Gott}}, \bibinfo {author}
  {\bibfnamefont {C.}~\bibnamefont {Hillairet}}, \bibinfo {author}
  {\bibfnamefont {S.}~\bibnamefont {Jaure}}, \bibinfo {author} {\bibfnamefont
  {A.}~\bibnamefont {Leblanc}}, \bibinfo {author} {\bibfnamefont
  {R.}~\bibnamefont {Lehe}}, \bibinfo {author} {\bibfnamefont {A.}~\bibnamefont
  {Myers}}, \bibinfo {author} {\bibfnamefont {C.}~\bibnamefont {Piechurski}},
  \bibinfo {author} {\bibfnamefont {M.}~\bibnamefont {Sato}}, \bibinfo {author}
  {\bibfnamefont {N.}~\bibnamefont {Zaim}}, \bibinfo {author} {\bibfnamefont
  {W.}~\bibnamefont {Zhang}}, \bibinfo {author} {\bibfnamefont
  {J.}~\bibnamefont {Vay}},\ and\ \bibinfo {author} {\bibfnamefont
  {H.}~\bibnamefont {Vincenti}},\ }\bibfield  {title} {\bibinfo {title}
  {Pushing the frontier in the design of laser-based electron accelerators with
  groundbreaking mesh-refined particle-in-cell simulations on exascale-class
  supercomputers},\ }in\ \href {https://doi.ieeecomputersociety.org/} {\emph
  {\bibinfo {booktitle} {2022 SC22: International Conference for High
  Performance Computing, Networking, Storage and Analysis (SC) (SC)}}}\
  (\bibinfo  {publisher} {IEEE Computer Society},\ \bibinfo {address} {Los
  Alamitos, CA, USA},\ \bibinfo {year} {2022})\ pp.\ \bibinfo {pages}
  {25--36}\BibitemShut {NoStop}%
\end{thebibliography}%

\end{document}

\title{Supplemental Material for  ``Stability of the Modulator in a Plasma-Modulated Plasma Accelerator''}% Force line breaks with \\

\author{J. J. van de Wetering}
\email{johannes.vandewetering@physics.ox.ac.uk}
\affiliation{John Adams Institute for Accelerator Science and Department of Physics, University of Oxford, Denys Wilkinson Building, Keble Road, Oxford OX1 3RH, United Kingdom}%
\author{S. M. Hooker}%
\affiliation{John Adams Institute for Accelerator Science and Department of Physics, University of Oxford, Denys Wilkinson Building, Keble Road, Oxford OX1 3RH, United Kingdom}%
\author{R. Walczak}%
\affiliation{John Adams Institute for Accelerator Science and Department of Physics, University of Oxford, Denys Wilkinson Building, Keble Road, Oxford OX1 3RH, United Kingdom}%
\affiliation{Somerville College, Woodstock Road, Oxford OX2 6HD, United  Kingdom}%

\date{\today}% It is always \today, today,
             %  but any date may be explicitly specified

\maketitle

\section{Paraxial Description of Pulses in Plasma Channels}

\subsection{The Envelope Model}

Considering only the high frequency fields associated with the laser pulse, from Amp\`{e}re's law we find that the normalized vector potential $\bm{a}= e\bm{A}/m_ec$ evolves according to the wave equation
\begin{align}\label{eq:ampere}
    \left[\frac{\partial^2}{\partial t^2}-c^2\Delta+\frac{\omega_p^2}{n_0\gamma}\left(n_0+\delta n\right)\right]\bm{a} &= 0
\end{align}
where $\gamma = (1+\bm{p}^2+\bm{a}^2)^{1/2}$, $\bm{p}=\gamma m_e\bm{v}_\text{hf}$ and $\bm{v}_\text{hf}$ represents the high frequency quiver velocity of electrons which ignores the low frequency bulk fluid velocity from a wakefield response, $n_0$ is the unperturbed pre-formed plasma channel density and $\delta n$ represents the change in density due to the wake driven by the laser pulse. We have also assumed that the electrostatic response is small compared to the transverse current. To construct an envelope model for the laser evolution, consider a pulse of the form
\begin{align}
    \bm{a} = \hat{e}_La(r,\theta,z,t)\exp[i(k_Lz-\omega_Lt)]
\end{align}
where the laser frequency and wavenumber in free space $\omega_L, k_L$ are constants and $\hat{e}_L$ is the polarization of the laser, which we will treat as linearly polarized throughout this paper. It is convenient to shift from the lab frame axial coordinate $z$ and time $t$ to co-moving coordinates $\xi=z-v_gt$, and $\tau=t$, where $v_g/c = (1-\omega_{p0}^2/\omega_L^2)^{1/2}$ is defined as the group velocity of electromagnetic plane waves in uniform plasma at the on-axis plasma channel density $n_{00}=n_0(r=0)$, $\omega_{p0}=\omega_p(r=0)$, which may differ from the group velocity of a tightly focused laser pulse \cite{SpotVG}. Substituting a pulse of this form into Eq. (\ref{eq:ampere}) in the weakly relativistic and linear wake limit yields the following expression for the evolution of the envelope $a(r,\theta,\xi,\tau)$
\begin{align}\label{eq:full_envelope}
    &\Big[-2i\omega_L\frac{\partial }{\partial\tau}-2v_g\frac{\partial^2}{\partial\xi\partial\tau}-c^2\Delta_\perp + \frac{\partial^2}{\partial\tau^2} \nonumber \\ 
    &-(c^2-v_g^2)\frac{\partial^2}{\partial\xi^2} + \frac{\omega_p^2}{n_0}\left(\delta n_0+\delta n-n_0|a|^2/4\right)\Big]a = 0
\end{align}
where $\Delta_\perp\equiv(1/r)(\partial/\partial r)(r\partial/\partial r)+(1/r^2)\partial^2/\partial\theta^2$ and $\delta n_0(r)=n_0(r)-n_{00}$. This PDE, coupled to a self-consistent wake solution for $\delta n$, fully describes the evolution of the laser envelope in the axisymmetric, weakly relativistic, quasi-static linear wakefield regime. 

\subsection{The Paraxial Approximation}

To further simplify Eq. (\ref{eq:full_envelope}) to the paraxial approximation, we assume that the discrepancy between the group velocity and speed of light in vacuum is negligible ($v_g\approx c$) and that the pulse envelope has a large enough longitudinal extent that the term $\partial^2/\partial\xi^2$ can be neglected. We also assume that the envelope evolution is slow relative to the carrier frequency so that the terms $v_g\partial^2/\partial\xi\partial\tau$ and $\partial^2/\partial\tau^2$ time derivative terms can be dropped. This yields a generalized nonlinear paraxial wave equation in an axisymmetric plasma channel $n_0(r)=n_{00}+\delta n_0(r)$
\begin{align}\label{eq:GNLS}
    &\left[\frac{i}{\omega_L}\frac{\partial}{\partial\tau} + \frac{c^2}{2\omega_L^2}\Delta_\perp\right] a = \nonumber \\ &\frac{\omega_p^2}{2\omega_L^2n_0}\left[\delta n_0(r) + \delta n(r,\xi;|a|^2) - n_0(r)|a|^2/4\right]a
\end{align}
where the nonlinearities come from relativistic effects and the interaction between the pulse and its own excited wake.

\subsection{Matched Plasma Channels}

Using Eq. (\ref{eq:GNLS}) we can derive the form of the matched plasma channel, which confines a Gaussian beam with a constant spot size as it propagates. Assuming no contributions from relativistic self-focusing nor from plasma wakes, this yields the linear paraxial wave equation in a channel 
\begin{align}
    \left[\frac{i}{\omega_L}\frac{\partial}{\partial\tau} + \frac{c^2}{2\omega_L^2}\Delta_\perp\right] a &= \frac{\omega_p^2\delta n_0(r)}{2\omega_L^2n_0}a\,.
\end{align}
Substituting a Gaussian pulse with a fixed spot size $w_0$ yields the following stable solution
\begin{align}\label{eq:gausschannel}
    &a(r,\xi,\tau) = a_0f(\xi)\exp\left(-\frac{r^2}{w_0^2}-i\omega_L\tau\frac{2c^2}{\omega_L^2w_0^2}\right)\,,\nonumber \\
    &n_0(r) = n_{00} + \Delta n(r/w_0)^2\,,\,\,\,\, \Delta n = (\pi r_ew_0^2)^{-1}
\end{align}
where the longitudinal profile of the pulse, $0\leq f(\xi)\leq1$, is assumed to be sufficiently slowly varying for the paraxial approximation to hold, $n_{00}$ is an arbitrary on-axis density and $r_e$ is the classical electron radius. This result for the matched channel can also be shown to confine all Laguerre-Gaussian modes with envelope solutions of the form
\begin{align}\label{eq:LGchannel}
    &a_{pm}(r,\theta,\xi,\tau) = \nonumber \\ &\alpha_{pm}(\xi)\exp\left(-i\omega_L\tau(2p+|m|+1)\frac{2c^2}{\omega_L^2w_0^2}\right)\text{LG}_{pm}\,, \nonumber \\
    &\text{LG}_{pm}(r,\theta) = \nonumber \\ &\sqrt{\frac{p!}{(p+|m|)!}}\left(\frac{\sqrt{2}\,r}{w_0}\right)^{|m|}\exp\left(-\frac{r^2}{w_0^2}+im\theta\right)L^{|m|}_p\left(\frac{2r^2}{w_0^2}\right)\,, \nonumber \\
    &\langle\text{LG}_{p'm'}|(\ldots)|\text{LG}_{pm}\rangle \equiv \nonumber \\
    &\frac{2}{\pi w_0^2}\int_0^{2\pi}d\theta\int_0^\infty rdr \text{LG}_{p'm'}^*(\ldots)\text{LG}_{pm}
\end{align}
where the integers $p\geq0$ and $m$ are the radial and azimuthal indexes respectively and the $L^{|m|}_p$ functions are the generalized Laguerre polynomials. Note that it is the interference between the fundamental and first azimuthal and radial modes that set the laser centroid and spot size oscillation frequencies $\omega_c=\omega_w/2=2c^2/\omega_Lw_0^2$ respectively.

\section{2D Slab vs 3D Cylindrical Geometry}

To justify the use of 2D PIC simulations to study the stability of the plasma modulator, we outline the pulse propagation theory in 2D slab geometry. The physical description of seeded spectral modulation in 2D slab and 3D cylindrical are similar, but they have some key differences which come from the transverse Laplacian operator $\Delta_\perp$ taking a different form. The matched parabolic channel for a Gaussian beam with spot size $w_0$ still takes the same form $n_0(x)=n_{00}+\Delta n(x/w_0)^2$, but the confined modes are instead described by Hermite-Gaussian functions
\begin{align}\label{eq:HGchannel}
    &a_l(x,\xi,\tau) = \alpha_l(\xi)\exp\left(-i\omega_L\tau(l+1)\frac{2c^2}{\omega_L^2w_0^2}\right)\text{HG}_l\,, \nonumber \\
    &\text{HG}_l(x) = \sqrt{\frac{1}{2^l l!}}\exp\left(-\frac{x^2}{w_0^2}\right)H_l\left(\frac{\sqrt{2}\,x}{w_0}\right)\,, \nonumber \\
    &\langle\text{HG}_{l'}|(...)|\text{HG}_l\rangle \equiv \sqrt{\frac{2}{\pi w_0^2}}\int_{-\infty}^\infty dx \text{HG}_{l'}^*(...)\text{HG}_l
\end{align}
where integer $l\geq 0$ is the transverse index and the $H_l$ functions are the (physicist's) Hermite polynomials. Note that the laser centroid and spot size oscillation frequencies are the same in both 2D slab and 3D cylindrical geometry, so we would expect processes tied to spot/centroid oscillations to behave similarly in 2D and 3D. However, the rate of spectral modulation does depend on the dimensionality. For example, consider the shallow channel limit $\Delta n\ll n_{00}$ where the seed wake can be written in the form
\begin{align}
    \delta n = \delta n_s\cos(k_{p0}\xi)
    \begin{cases}
        1 & \text{1D}\\
        e^{-2x^2/w_0^2} & \text{2D}\\
        e^{-2r^2/w_0^2} & \text{3D}
    \end{cases} 
\end{align}
where $\delta n_s$ denotes the on-axis seed wake amplitude. Assuming that the modulating drive pulse remains in the fundamental channel mode, the spectral modulation rate parameter $\Omega_s$ is given by
\begin{align}
    \Omega_s = \frac{1}{4}\frac{\omega_{p0}^2}{\omega_L}\frac{\delta n_s}{n_{00}}
    \begin{cases}
         1 & \text{1D}\\
        \langle\text{HG}_0|e^{-2x^2/w_0^2}|\text{HG}_0\rangle = 1/\sqrt{2} & \text{2D}\\
        \langle\text{LG}_{00}|e^{-2r^2/w_0^2}|\text{LG}_{00}\rangle = 1/2 & \text{3D}
    \end{cases} 
\end{align}
hence 2D PIC simulations are expected to spectrally modulate $\sim\sqrt{2}$ times faster than predicted by 3D cylindrical theory (and $\sim\sqrt{2}$ times slower than 1D theory).

The suppression of spectral modulation by wake phase-front curvature caused by the plasma channel, as described in the paper, also changes depending on the dimensionality. This is again primarily caused by the difference between $\langle\text{HG}_0|\delta n(x,\xi)|\text{HG}_0\rangle$ and $\langle\text{LG}_{00}|\delta n(r,\xi)|\text{LG}_{00}\rangle$, resulting in the suppression towards the pulse tail being more pronounced in the 3D cylindrical case. There is also a smaller effect due to differences in the wake structure itself between 2D and 3D.

\section{Spectral Phase for Pulse Train Formation}

As derived previously in 1D by Jakobsson et al \cite{PhysRevLett.127.184801}, to first order the spectral modulation takes the form of a set of sidebands of angular frequencies $\omega_m = \omega_L + m \omega_{p0}$, where $m = \pm1, \pm2, \pm3, \ldots$ is the sideband order. These sidebands will each have a relative spectral phase $\psi_m = -|m|\pi/2$. With a more detailed analysis, which can be derived from Eq. \eqref{eq:TDPT_nonparaxial_2level} assuming a narrow bandwidth $\tau_\text{drive}^{-1}\ll \omega_{p0}$ and that light only remains in the fundamental channel mode, the full relative spectral phase of the modulated drive pulse before compression into a pulse train can be approximately described by the following nearest integer staircase function
\begin{align}
    \psi(\omega) = -\left|\text{nint}\left(\frac{\omega-\omega_L}{\omega_{p0}}\right)\right|\frac{\pi}{2}\,.
\end{align}
To form a pulse train, we wish to remove this spectral phase from the pulse. However, it is not practical to remove a spectral phase of this form with a dispersive optic. Instead, we can take advantage of the narrow bandwidth of each of the sidebands to approximately remove this spectral phase by applying a continuous dispersion function of the form
\begin{align}
    \psi^\text{opt}(\omega) = +\left|\frac{\omega-\omega_L}{\omega_{p0}}\right|\frac{\pi}{2}.
\end{align}
This form of $\psi^\text{opt}(\omega)$ was used in the main paper to evaluate the pulse trains that can be formed from the spectrally modulated drive pulses.

\section{Non-Paraxial Description of Seeded Spectral Modulation in Plasma Channels}

Unlike the paraxial equation, this description will include group velocity dispersion as well as asymmetries between the dynamics of the generated Stokes and anti-Stokes sidebands in the plasma modulator. We will start by following the procedure outlined by \cite{SpotVG} for deriving the non-paraxial description of pulses in fully ionized plasma, but here we will include contributions from a parabolic plasma channel and a seed wake. We begin with the full 3D wave equation for a laser pulse propagating in a fully ionized plasma
\begin{align}
    \left(\Delta-\frac{1}{c^2}\frac{\partial^2}{\partial t^2}-k_p^2\right)\bm{a} = 0
\end{align}
where $\bm{a}=e\bm{A}/m_ec$ is the normalized vector potential and we have neglected higher order plasma source terms from having a finite electrostatic potential \cite{SpotVGplasma}. We now switch to new coordinates $\xi=z-ct$, $\eta=(z+ct)/2$, which yields
\begin{align}
    \left(2\frac{\partial^2}{\partial\xi\partial\eta}+\Delta_\perp-k_p^2\right)\bm{a} = 0\,.
\end{align}
We seek envelope solutions in the form $\bm{a}=\left[ a\exp(ik_L\xi)+\text{c.c.}\right]\hat{e}_L/2$ where $k_L$ is a constant. This yields the envelope PDE
\begin{align}
    \left[2\left(ik_L+\frac{\partial}{\partial\xi}\right)\frac{\partial}{\partial\eta}+\Delta_\perp-k_p^2\right] a(r,\xi,\eta) = 0\,.
\end{align}
We then take the Fourier transform in the variable $\xi$ and apply the convolution theorem
\begin{align}
    &\left[2i\left(k_L+k\right)\frac{\partial}{\partial\eta}+\Delta_\perp\right] a_k = \left(k_p^2\right)_k\ast a_k\,, \nonumber \\
    & a_k(r,k,\eta) = \frac{1}{\sqrt{2\pi}}\int_{-\infty}^\infty d\xi e^{-ik\xi} a(r,\xi,\eta)\,,\nonumber \\
    &(f\ast g)(k) := \frac{1}{\sqrt{2\pi}}\int_{-\infty}^\infty f(k')g(k-k')dk'
\end{align}
where we have included a $1/\sqrt{2\pi}$ normalization in the definition of the convolution for notational convenience. Ignoring relativistic effects, we can split $k_p^2(r,\xi,\eta)$ into contributions from a pre-formed axisymmetric plasma channel $n_0(r)$ and a plasma wake $\delta n(r,\xi,\eta)$
\begin{align}
    &\left[i(1+k/k_L)\frac{1}{k_L}\frac{\partial}{\partial\eta}+\frac{1}{2k_L^2}\Delta_\perp-\frac{2}{k_L^2w_0^2}\frac{n_0(r)}{\Delta n}\right] a_k = \nonumber \\
    &\frac{2}{k_L^2w_0^2}\left(\frac{\delta n(r,\xi,\eta)}{\Delta n}\right)_k\ast a_k
\end{align}
where $\Delta n \equiv (\pi r_e w_0^2)^{-1}$ is the channel depth parameter. Ignoring the wake for now, a matched parabolic channel $n_0(r)=n_{00}+\Delta n(r/w_0)^2$ can guide any linear combination of Laguerre-Gaussian modes of the form
\begin{align}
    & a_k^{pm}(r,\theta,k,\eta) =   \alpha_k^{pm}(k)\exp[-i\tilde{k}_k^{pm}(k)\eta]\text{LG}_{pm}(r,\theta)\,, \nonumber \\
    &\tilde{k}_k^{pm}(k) = \frac{2}{k_Lw_0^2}\frac{2p+|m|+1+n_{00}/\Delta n}{1+k/k_L}\,, \nonumber \\
    &\text{LG}_{pm}(r,\theta) = \nonumber \\
    &\sqrt{\frac{p!}{(p+|m|)!}}\left(\frac{\sqrt{2}\,r}{w_0}\right)^{|m|}\exp\left(-\frac{r^2}{w_0^2}+im\theta\right)L^{|m|}_p\left(\frac{2r^2}{w_0^2}\right)\,.    
\end{align}
Assuming that $k/k_L\ll 1$ remains valid for the majority of the pulse $a_k$, (i.e. for pulse durations that are not too short relative to the laser cycle period), we can also write the Laguerre-Gaussian mode solutions in real space to first order in the form
\begin{align}
     a^{pm}(r,\theta,\xi,\eta) &=   \alpha^{pm}(\xi)\ast\left(\exp[-i\tilde{k}_k^{pm}(k)\eta]\right)_\xi\text{LG}_{pm}(r,\theta) \nonumber \\
    &\approx   \alpha^{pm}(\xi+\tilde{k}_0^{pm}\eta/k_L)\exp(-i\tilde{k}_0^{pm}\eta)\text{LG}_{pm}(r,\theta)\,, \nonumber \\
    \tilde{k}_0^{pm}/k_L &= \frac{\omega_{p0}^2}{2k_L^2c^2}+(2p+|m|+1)\frac{2}{k_L^2w_0^2}
\end{align}
where $\alpha^{pm}(\xi)$ is the inverse Fourier transform of $\alpha^{pm}_k(k)$. This describes the first order group velocity dispersion and wavenumbers of Laguerre-Gaussian modes due to the on-axis plasma density and finite spot size effects. We can see that for a pulse primarily in the fundamental mode, after every spot size oscillation the first radial mode will fall behind the fundamental mode by a laser wavelength (and similarly for centroid oscillations). Hence we eventually need to take this group velocity dispersion into account if the pulse propagates over many spot size oscillations in a long plasma channel.

If we can treat the wake contribution as a small perturbation to the matched parabolic plasma channel, we can use time-dependent perturbation theory to calculate the transitions between the channel modes with the following expression
\begin{widetext}
\begin{align}\label{eq:TDPT}
    &i(1+k/k_L)\frac{\partial  \alpha_k^{pm}(k,\eta)}{\partial\eta}\exp(-i\tilde{k}_k^{pm}\eta) = k_c\sum_{p'm'}\Big\langle\text{LG}_{pm}\Big|\left(\frac{\delta n(r,\xi,\eta)}{\Delta n}\right)_k\ast\left(  \alpha_k^{p'm'}(k,\eta)\exp(-i\tilde{k}_k^{p'm'}\eta)\right)\Big|\text{LG}_{p'm'}\Big\rangle
\end{align}
\end{widetext}
where $k_c=k_w/2=2/k_Lw_0^2$ are the centroid and spot size oscillation wavenumbers respectively.

Assume that the short seed pulse has the same wavelength as the drive pulse, has many laser cycles in its duration, is in the fundamental mode and does not appreciably deplete. The seed pulse will then have a group velocity of
\begin{align}
    v_{g,s}/c = 1 - \tilde{k}_0^{00}/k_L = 1 - \frac{\omega_{p0}^2}{2k_L^2c^2}-\frac{2}{k_L^2w_0^2}\,.
\end{align}
Note that the group velocity is slowed by both plasma and finite spot size effects. This means that in general the seed wake will be in the form
\begin{align}
    \delta n(r,\xi,\eta) = \delta n(r,\xi+\tilde{k}_0^{00}\eta/k_L)\,.
\end{align}
We now choose to work in the shallow channel limit $\Delta n \ll n_{00}$ to ignore the non-separable transverse wake structure introduced by the channel \cite{Andreev96}. Note that using a square-like channel would achieve a similar effect, but would have Bessel modes rather than Laguerre-Gaussian modes. The wake excited by the seed pulse considering both plasma and finite spot size effects on its group velocity in this limit takes the form
\begin{align}
    \delta n(r,\xi,\eta) = \delta n_s\text{LG}_{00}^2(r)\cos\left[k_{p0}\left(\xi+\tilde{k}_0^{00}\eta/k_L\right)\right]\,.
\end{align}
Substituting this seed wake into Eq. \eqref{eq:TDPT} yields the non-paraxial plasma modulator equation
\begin{widetext}
\begin{align}\label{eq:TDPT_nonparaxial}
    &i\frac{1+k/k_L}{k_L}\frac{\partial  \alpha_k^{pm}}{\partial\eta} = \frac{1}{4}\frac{\omega_{p0}^2}{k_L^2c^2}\frac{\delta n_s}{n_{00}}\sum_{p'm'}  \alpha^{p'm'}_{k\pm k_{p0}}\exp\left[-i\left(\tilde{k}_{k\pm k_{p0}}^{p'm'}-\tilde{k}_k^{pm}\pm(k_{p0}/k_L)\tilde{k}_0^{00}\right)\eta\right]\langle\text{LG}_{pm}|\text{LG}_{00}^2|\text{LG}_{p'm'}\rangle
\end{align}
where we can now clearly see that the seed wake modulating a pulse with an initially short bandwidth will generate Stokes and anti-Stokes sidebands in $k$-space separated by integer multiples of $k_{p0}$. Assuming that $k,k_{p0}\ll k_L$, the wavenumber shifts approximate to
\begin{align}
    &\left(\tilde{k}_{k\pm k_{p0}}^{p'm'}-\tilde{k}_k^{pm}\pm(k_{p0}/k_L)\tilde{k}_0^{00}\right)/k_c \approx 
    \left[2(p'-p)+(|m'|-|m|)\right](1-k/k_L) \mp \left(2p'+|m'|\right)(k_{p0}/k_L)\,.
\end{align}
Using this expression and assuming that most of the light remains in the fundamental mode and that no azimuthal modes are present, we can approximate the plasma modulator equation as a two-level system of the fundamental and first radial modes
\begin{align}\label{eq:TDPT_nonparaxial_2level}
    &i\frac{1+k/k_L}{k_L}\frac{\partial  \alpha_k^{00}}{\partial\eta} = \frac{1}{8}\frac{\omega_{p0}^2}{k_L^2c^2}\frac{\delta n_s}{n_{00}}\left(  \alpha^{00}_{k\pm k_{p0}}
    + \tfrac{1}{2}  \alpha^{10}_{k\pm k_{p0}}\exp\left[-ik_w\left(1-k/k_L\mp k_{p0}/k_L\right)\eta\right]\right)\,, \nonumber \\
    &i\frac{1+k/k_L}{k_L}\frac{\partial  \alpha_k^{10}}{\partial\eta} = \frac{1}{8}\frac{\omega_{p0}^2}{k_L^2c^2}\frac{\delta n_s}{n_{00}}\left(\tfrac{1}{2}  \alpha^{00}_{k\pm k_{p0}}\exp\left[ik_w(1-k/k_L)\eta\right]
    + \tfrac{1}{2}  \alpha^{10}_{k\pm k_{p0}}\exp\left[\pm ik_w\left(k_{p0}/k_L\right)\eta\right]\right)
\end{align}
\end{widetext}
which to zeroth order in $k/k_L$ and $k_{p0}/k_L$ gives the same result given by the paraxial equation used in the paper (apart from the slightly different propagation variable $\eta$). However, to first order we see symmetry-breaking between the Stokes and anti-Stokes sidebands which was not captured by the paraxial equation. We see here that the Stokes sidebands are generated faster by the seed wake and also undergo faster spot size oscillations and transverse mode transitions than the anti-Stokes. This asymmetry in the dynamics between the Stokes and anti-Stokes sidebands is necessary to explain the transverse separation of Stokes and anti-Stokes light observed in PIC simulations in the regime where the self-wake of the modulating drive pulse is no longer negligible.

\section{Particle-in-Cell Simulations}

Two-dimensional simulations were performed with the PIC code WarpX (version 22.07) \cite{WarpX}. Results from eight simulations are included in the paper with laser-plasma parameters and respective figures outlined in Table \ref{simParams}. All of these eight simulations were performed at an on-axis density of $n_{00}=\SI{2.5e17}{cm^{-3}}$ in a modulator of length $L_\text{mod}=\SI{110}{mm}$ with seed and drive pulses with the same wavelength $\lambda_L=\SI{1030}{nm}$ and spot size $w_0=$ 30 or 50 $\SI{}{\micro m}$. The seed and drive pulses were polarized out of and in the plane of simulation respectively. All simulations had a longitudinal resolution of $\Delta z = \lambda_L/50$ and transverse resolution of $\Delta x = \lambda_L/2.5$ using a second order Yee field solver and perfectly matched layer (PML) transverse boundary conditions. The transverse window size for all simulations was at least $2.67w_0$ away from the axis. The ``square'' and ``parabolic'' plasma channels were parameterized in the following form
\begin{align}
    &\frac{n_0^\text{square}(r)-n_{00}}{\Delta n} =  \nonumber \\
    &\begin{cases} 
        (r/w_0)^{10} & 0 \leq r < 1.2w_0 \\ 
        1.2^{10} & 1.2w_0 \leq r < 1.2w_0 + d \\ 
        1.2^{10}\left(1-\frac{r-1.2w_0-d}{d}\right) & 1.2w_0 + d \leq r < 1.2w_0 + 2d \\
        0 & r \geq 1.2w_0 + 2d
    \end{cases} \nonumber \\
    &\frac{n_0^\text{parabolic}(r)-n_{00}}{\Delta n} =  \nonumber \\
    &\begin{cases} 
        (r/w_0)^{2} & 0 \leq r < 2w_0 \\ 
        4 & 2w_0 \leq r < 2w_0 + d \\ 
        4\left(1-\frac{r-2w_0-d}{d}\right) & 2w_0 + d \leq r < 2w_0 + 2 d \\
        0 & r \geq 2 w_0 + 2 d
    \end{cases} \nonumber \\
\end{align}
where $d=\SI{10}{\micro m}$. Note that all simulations were initialized with a laser pulse with a gaussian transverse profile of spot size $w_0$, which differs slightly from the fundamental guiding mode of the square channel.

\begin{widetext}
\begin{center}
\begin{table}[!h]
\begin{tabular}{ |p{1.8cm}||p{1.6cm}|p{1.4cm}|p{1.8cm}|p{1.8cm}|p{1.6cm}|p{1.6cm}|p{1.8cm}|  }
 \hline
 \multicolumn{8}{|c|}{PIC Simulation Parameters} \\
 \hline
 Simulation & Figures & $w_0$ ($\SI{}{\micro m}$) & $W_\text{seed}$ (mJ) & $W_\text{drive}$ (J) & $\tau_\text{seed}$ (fs) & $\tau_\text{drive}$ (ps) & Channel\\
 \hline
 $(i)$ & 2, 3 & 30 & 50 & 0.6 & 40 & 1.0 & parabolic \\
 $(ii)$ & 3, 4a & 30 & 50 & 0.6 & 40 & 1.0 & square \\
 $(iii)$ & 3 & 50 & 139 & 1.67 & 40 & 1.0 & parabolic \\
 $(iv)$ & 4b, 5b & 30 & 50 & 1.2 & 40 & 1.0 & square \\
 $(v)$ & 4c & 30 & 50 & 2.4 & 40 & 4.0 & square \\
 $(vi)$ & 5a & 30 & 50 & 1.2 & 40 & 0.25 & square \\
 $(vii)$ & 5c & 30 & 50 & 1.2 & 40 & 2.0 & square \\
 $(viii)$ & 5d & 30 & 50 & 1.2 & 40 & 4.0 & square \\
 \hline
\end{tabular}
\caption{\label{simParams}}
\end{table}
\end{center}
\end{widetext}

\acknowledgements

This research was funded in whole, or in part, by EPSRC and STFC, which are Plan S funders. For the purpose of Open Access, the author has applied a CC BY public copyright licence to any Author Accepted Manuscript version arising from this submission.

\bibliography{references_supp}% Produces the bibliography via BibTeX.